\documentclass[aps,reprint, amsmath,amssymb,aps]{revtex4-1}
\usepackage{graphicx}
\usepackage{dcolumn}
\usepackage{bm}
\usepackage{mathrsfs}
\usepackage{hyperref}
\allowdisplaybreaks[4]

\begin{document}
\title{Tunable surface configuration of skyrmions in cubic helimagnets}
\author{Xuejin Wan}
\author{Yangfan Hu}
 \email[Corresponding author.]{huyf3@mail.sysu.edu.cn}
\author{Biao Wang}
 \email[Corresponding author.]{wangbiao@mail.sysu.edu.cn}
\affiliation{Sino-French Institute of Nuclear Engineering and Technology,Sun Yat-sen University,510275 GZ, China}

\begin{abstract}
In bulk helimagnets, the presence of magnetic skyrmion lattices is always accompanied by a periodic stress field due to the intrinsic magnetoelastic coupling. The release of this nontrivial stress field at the surface causes a periodic displacement field, which characterizes a novel particle-like property of skyrmion: its surface configuration. Here, we derive the analytical solution of this displacement field for semi-infinite cubic helimagnets when skyrmions are present. For MnSi, we show that the skyrmion lattices have a bumpy surface configuration characterized by periodically arranged peaks with a characteristic height of about 10$^{-13}$ m. The pattern of the peaks can be controlled by varying the strength of the applied magnetic field. Moreover, we prove that the surface configuration varies together with the motion and deformation of the skyrmion lattices. As a result, the surface configuration can be tuned by application of electric current, mechanical loads, as well as any other form of external field which has an effect on the skyrmions.
\end{abstract} 
\maketitle

\section{Introduction}
A magnetic skyrmion is a topologically nontrivial spin texture. Periodic arrangement of skyrmions can be approximated by a superposition of three single-Q helices whose wavevectors form an equilateral triangle, and is thus referred to as a triple-Q structure \cite{1,2,59}. The existence of skyrmions in helimagnets has been theoretically predicted several decades ago \cite{3,4}, while the first successful experimental observation was achieved in cubic helimagnet MnSi in 2009 \cite{1}. Later, other helimagnets which can host skyrmions were found, such as FeGe \cite{5}, Fe$_x$Co$_{1-x}$Si \cite{6} and Mn$_{1-x}$Fe$_x$Si \cite{7}. In noncentrosymmetric helimagnets, due to the spin-orbit coupling and the lack of inverse symmetry, Dzyaloshinskii-Moriya (DM) interaction arises \cite{8}. Under an appropriate applied magnetic field, the competition between DM energy, favoring spin rotations, and ferromagnetic exchange energy, favoring spin alignment, induces the intriguing skyrmion phase \cite{9}. As a magnetic phase, skyrmions have great potential in the next-generation magnetic storage devices because of their small size, facile current-driven motion \cite{10}, and particle-like nature \cite{11,12}.

Magnetic skyrmions share many properties with single particles. They are localized in space and have a long lifetime. They are topologically protected \cite{51,52}, in the sense that the topological integer characterizing them is 1, different from other magnetic structures with topological integer 0, such as helical phase and ferromagnetic phase. They give rise to elementary excitations with rotational mode and breathing mode \cite{15}. Moreover, The system hosting skyrmions may undergo a phase transition from skyrmion phase to skyrmion glass structure \cite{14}. Here we would like to discuss another particle-like property of magnetic skyrmions: their surface configuration.

In helimagnets, interaction between the elastic field and the skyrmion phase due to magnetoelastic coupling occurs in two different energy scales. The strong one is phase-transition-related, for instance, the creation and annihilation of skyrmions in MnSi by uniaxial stress \cite{16,17} and the jump of elastic stiffnesses $C_{11}$ and $C_{33}$ of MnSi \cite{18}. The weak one is related to the elastic property of the skyrmion phase , for example, the emergent deformation of skyrmion lattices in FeGe induced by anisotropic strain \cite{19} and the periodic elastic field accompanying magnetic skyrmions \cite{20}. For semi-infinite helimagnets with magnetoelastic coupling, the incompatibility between the skyrmion-induced periodic stress field and the free surface boundary condition will inevitably lead to a displacement field, suggesting that the surface configuration of the material is altered due to the presence of skyrmions.

In this paper, we derive the analytical solution of displacement field for semi-infinite cubic helimagnets hosting skyrmions. Due to magnetoelastic coupling, the peculiar magnetic structure of skyrmions will induce incompatible eigenstrains and further lead to eigenstresses. At the surface, to meet the stress-free requirement, a fictitious force distribution $F$ is applied to balance the eigenstresses, which causes a surface-induced displacement field. Therefore, the total displacement field for semi-infinite cubic helimagnets hosting skyrmions is composed of a skyrmion-induced displacement field and a surface-induced displacement field. The former part has been derived in one of our previous work \cite{20}, and the latter part is to be solved here. The fictitious force distribution can be decomposed into two kinds: one results in 2D plane strain problems and the other one results in 3D elastic problems. The general elastic solution for the 2D plane strain problem is derived by using the Airy stress function and Fourier transform, where the functional regularization method is used to treat the non-convergence issue of the integral form of displacement.The 3D problem can be easily solved due to the harmonic form of the force distribution. The analytical displacement field is finally obtained by substituting $F$ into the 2D and 3D solutions. For MnSi, the normal displacement field is found to be dominated by two triple-Q structures $u^{s1}_3$ and $u^{s3}_3$. $u^{s1}_3$ undergoes a ``configurational reversal'' and $u^{s3}_3$ remains almost unchanged when the external magnetic field increases from 0.1 T to 0.4 T, resulting in varying surface configuration characterized by periodically arranged peaks. We further demonstrate that the surface displacement field moves or deforms with the motion or deformation of skyrmion lattices. Hence, we have proved the tunability of this displacement field, and that it characterizes the shape of skyrmion lattices at the surface under various kinds of applied field.

\section{Elasticity problem for semi-infinite cubic helimagnets in the skyrmion phase with a free surface}
Following the unified theory of magnetoelastic effects in B20 compounds developed in Ref. \cite{20}, we write the Helmholtz free energy density for cubic helimagnets in the form:
\begin{equation}
\begin{aligned}
w=&\sum_{i=1}^3A(\frac{\partial\bm M} {\partial {x_i}})^2-\bm B\bm\cdot\bm M+b\bm M\bm\cdot (\bm\nabla \times \bm M)
\\&+w_{an}+w_L+w_{el}+w_{me},
\end{aligned}
\label{1}
\end{equation} where the first three terms represent respectively the Heisenberg exchange energy density with stiffness $A$, the Zeeman energy density with external applied magnetic field $\bm B$ and the DM interaction with Dzyaloshinskii constant $b$; $w_{an}=\sum_{i=1}^3B_cM_i^4$ is cubic anisotropy term; $w_L=\alpha_1 (T-T_0)\bm M^2+\alpha_2 \bm M^4$ includes two Landau expansion terms. The last two terms in Eq. (\ref{1}) are related to the strains. $w_{el}$ is the elastic energy density and $w_{me}$ the magnetoelastic energy density,
\begin{equation}
\begin{aligned}
w_{el}=&\frac 1 2C_{11}(\varepsilon_{11}^2+\varepsilon_{22}^2+\varepsilon_{33}^2)+C_{12}(\varepsilon_{11}\varepsilon_{22}+
\\&\varepsilon_{11}\varepsilon_{33}+\varepsilon_{22}\varepsilon_{33})+\frac 1 2C_{44}(\gamma_{12}^2+\gamma_{13}^2+\gamma_{23}^2),
\label{2}
\end{aligned}
\end{equation}
\begin{equation}
\begin{aligned}
w_{me}=&\frac 1{M^2_s}[L_1(M^2_1\varepsilon_{11}+M^2_2\varepsilon_{22}+M^2_3\varepsilon_{33})
\\&+L_2(M^2_3\varepsilon_{11}+M^2_1\varepsilon_{22}+M^2_2\varepsilon_{33})
\\&+L_3(M_1M_2\gamma_{12}+M_1M_3\gamma_{13}+M_2M_3\gamma_{23})
\\&+KM^2\varepsilon_{ii}+\sum^6_{i=1}L_{Oi}f_{Oi}],
\label{3}
\end{aligned}
\end{equation}
where $\gamma_{ij}=2\varepsilon_{ij} \;(i,j=1,2,3 \text{ and } i\neq j)$ are the engineering shear strains, $\varepsilon_{ij} \;(i,j=1,2,3)$ are the strains, $C_{11}$, $C_{22}$ and $C_{44}$ are the elastic constants for cubic crystals, $M_s$ is the saturation magnetization, $M_i\;(i=1,2,3)$ are the magnetization components satisfying $M^2=M_1^2+M_2^2+M_3^2$, $L_i \; (i=1,2,3)$ and $L_{Oi}\;  (i=1,...,6)$ are  magnetoelastic coupling constants and $f_{Oi}\;  (i=1,...,6)$ represent high order magnetoelastic coupling terms whose detailed expressions are given in Ref. \cite{20}.

In the conventional Cartesian coordinate system O-XYZ for cubic crystals, where the cartesian axes $X$, $Y$ and $Z$ are collinear with the crystallographic axes $a$, $b$ and $c$, respectively, the triple-Q structure of magnetization field for skyrmion phase stabilized by applied magnetic field along [0 0 1] direction can be described as the form:
\begin{equation}
\begin{aligned}
\bm M=&\begin{bmatrix} 0\\0\\M\mathrm{cos}(\varphi) \end{bmatrix}+\frac {\sqrt 3M\mathrm{sin}(\varphi)}3\left \{\begin{bmatrix} 0\\\mathrm{sin}(\bm q_1\bm r)\\-\mathrm{cos}(\bm q_1\bm r) \end{bmatrix} \right .
\\& \left .+\begin{bmatrix} -\frac {\sqrt3} 2  \mathrm{sin}(\bm q_2\bm r)\\ -\frac 12 \mathrm{sin}(\bm q_2\bm r)\\-\mathrm{cos}(\bm q_2\bm r)\end{bmatrix}
+\begin{bmatrix} \frac {\sqrt3} 2  \mathrm{sin}(\bm q_3\bm r)\\ -\frac 12 \mathrm{sin}(\bm q_3\bm r)\\-\mathrm{cos}(\bm q_3\bm r) \end{bmatrix}\right\},
\label{5}
\end{aligned}
\end{equation}
where $\varphi$ is the angle between magnetization vector and $Z$-axis, $\bm q_1=q[1,0,0]^T,\; \bm q_2=q[-\frac 12,\frac {\sqrt 3}2,0]^T, \; \bm q_3=-\bm q_1-\bm q_2$ are wavevectors with magnitude $q$, and $ \bm r$ is the Cartesian coordinate.

For a bulk cubic crystal free from body forces and surface constrains, the incompatible eigenstrains induced by skyrmions leads to eigenstresses. In a semi-infinitely extended material (illustrated in FIG. 1) with eigenstresses induced by skyrmions, to set the surface boundary $z=0$ stress-free, equal and opposite surface force should be applied, the force needed has the components
\begin{equation}
\begin{bmatrix} F_1\\F_2\\F_3 \end{bmatrix}=-\begin{bmatrix} \sigma_{11}\; \sigma_{12}\; \sigma_{13}\\ \sigma_{21}\; \sigma_{22} \; \sigma_{23}\\ \sigma_{31}\; \sigma_{32} \;\sigma_{33} \end{bmatrix}\begin{bmatrix} 0\\0\\1 \end{bmatrix}=-\begin{bmatrix} \sigma_{13}\\ \sigma_{23}\\ \sigma_{33} \end{bmatrix}.
\label{16}
\end{equation}
Due to the superposition of three triple-Q structures of the elastic field, $\sigma_{i3}\; (i=1,2,3)$ can be expressed in the following form
\begin{equation}
\sigma_{i3}=\sum_{j=1}^3\sigma_{i3}^{Sj}
\label{17},
\end{equation}
where the analytical expressions of the eigenstress components $\sigma_{i3}^{Sj},\;(i,j=1,2,3)$ are derived as Eqs. (\ref{7}-\ref{15}) in the appendix.

We would like to stress that $\sigma_{33}^{S1}$, whose sign is determined by $\varphi$ (the angle  between magnetization vector and $z$-axis) , undergoes a ``configurational reversal'' \cite{20}; while, $\sigma_{33}^{S3}$, which is linear with respect to $\mathrm{sin}^2(\varphi)$, is almost constant when the applied magnetic field changes.

\section{Two-dimensional half space elastic problem of cubic crystals}

\begin{figure}
\centering
\includegraphics[scale=0.7]{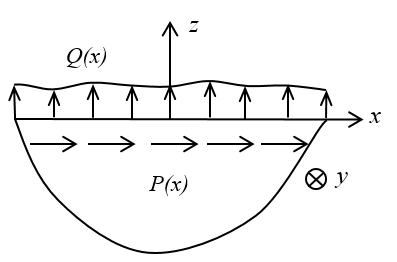}
\caption{Semi-infinitely extended cubic crystal subjected to surface normal force $Q(x)$ and surface shear force $P(x)$}
\end{figure}

Consider a semi-infinite domain defined by $z\leq 0$ illustrated in FIG. 1, where O-xyz system is generated by rotating O-XYZ system around $Z$-axis with $\theta$; $Q(x)$ and $P(x)$ represent respectively the normal and the shear force distributions on the surface $z=0$. For $Q$-induced 2D plane strain problem,we introduce the Airy stress function $U$ so that
\begin{equation}
\begin{aligned}
&\sigma_{11}=U_{,33},
\\&\sigma_{33}=U_{,11},
\\&\sigma_{13}=-U_{,13},
\label{18}
\end{aligned}
\end{equation}
where $\sigma_{ij}$ are stresses and $U_{,ij}=\frac{\partial^2 U}{\partial x_i \partial x_j}$. The boundary condition can be then expressed as
\begin{equation}
(\sigma_{33})_{z=0}=(U_{,11})_{z=0}=Q(x).
\label{19}
\end{equation}
By combining Hooke's law for cubic crystals, equation of compatibility $\varepsilon_{11,33}+\varepsilon_{33,11}=2\varepsilon_{13,13}$, and formulae (\ref{18}), we can derive
\begin{equation}
\beta^4U_{,1111}+2\mu U_{,1133}+U_{,3333}=0.
\label{20}
\end{equation}
Here, $\mu$ and $\beta$ are parameters related to the rotation angle $\theta$ and the elastic coefficients. Applying Fourier transform $\mathscr{F}$, defined as $\mathscr{X}(\lambda,z)=\mathscr{F}(X(x,z))= \frac1{\sqrt2\pi} \int_{-\infty}^{+\infty}X(x,z)e^{i\lambda x}dx$, to compatibility condition (\ref{20}) and boundary condition (\ref{19}), we have
\begin{gather}
\mathscr{U}_{,3333}-2\mu' \lambda'^2 \mathscr{U}_{,33}+\lambda'^4 \mathscr{U}=0,
\label{21}
\\-\lambda^2(\mathscr{U})_{z=0}=\mathscr{Q}(\lambda),
\label{22}
\end{gather}
where $\mathscr{U}$ and $\mathscr{Q}$ are the Fourier integral forms of $U$ and $Q$ respectively; $\mu'=\frac{\mu}{\beta^2}$ and $\lambda'=\beta\lambda$. According to the boundedness condition of $\mathscr{U}$ and the boundary condition (\ref{22}), one arrives at
\begin{equation}
\mathscr{U}=\frac{\mathscr{Q(\lambda)}}{\lambda^2(t_1-t_2)}(t_2\mathrm e^{t_1|\lambda'|z}-t_1\mathrm e^{t_2|\lambda'|z}),
\label{23}
\end{equation}
where $t_1=\sqrt{\frac{(1+\mu')}2}+\sqrt{\frac{(1-\mu')}2}\mathrm i,\; t_2=\sqrt{\frac{(1+\mu')}2}- \sqrt{\frac{(1-\mu')}2}\mathrm i$.
By applying the convolution theorem to Fourier integral form of stresses, we obtain
\begin{equation}
\begin{aligned}
&\sigma_{11}=\frac 1{\pi}\int_{-\infty}^{+\infty}\frac{-\beta^2\sqrt{2(1+\mu')}z'(x-\xi)^2Q(\xi)}{(x-\xi)^4+z'^4+2(x-\xi)^2z'^2\mu'}d\xi,
\\&\sigma_{33}=\frac1{\pi}\int_{-\infty}^{+\infty}\frac{-\sqrt{2(1+\mu')}z'^3Q(\xi)}{(x-\xi)^4+z'^4+2(x-\xi)^2z'^2\mu'}d\xi,
\\&\sigma_{13}=\frac1{\pi}\int_{-\infty}^{+\infty}\frac{-\beta\sqrt{2(1+\mu')}z'^2(x-\xi)Q(\xi)}{(x-\xi)^4+z'^4+2(x-\xi)^2z'^2\mu'}d\xi,
\label{25}
\end{aligned}
\end{equation}
where $z'=\beta z$. For isotropic materials and $\theta=0$, we have $\beta=\mu=1$, the solution for stresses (\ref{25}) can be found in Ref. \cite{24}.

The Green's function method, which requires firstly $Q=\delta_0$ with $\delta$ the Dirac Delta function, is used to derive the solution of displacement field caused by an arbitrary $Q(x)$. The relation between displacements and stresses is obtained from Hook's law and geometric equations $\varepsilon_{ij}=\frac{u_{i,j}+u_{j,i}}2$,
\begin{gather}
\begin{aligned}
&u_{1,1}=S_{11}\sigma_{11}+S_{13}\sigma_{33},
\\&u_{3,3}=S_{31}\sigma_{33}+S_{33}\sigma_{33},
\end{aligned}
\label{27}
\\u_{1,3}+u_{3,1}=S_{55}\sigma_{13}.
\label{28}
\end{gather}
Here, $S_{11}=\frac{C_{33}}{C_{11}C_{33}-C_{13}^2} $,  $S_{13}=S_{31}=-\frac{C_{13}}{C_{11}C_{33}-C_{13}^2} $, $S_{33}=\frac{C_{11}}{C_{11}C_{33}-C_{13}^2} $ and $S_{55}=\frac{1}{C_{44}} $ with $C_{ij}$ the elastic coefficients in O-xyz system; the stresses are obtained by applying $Q=\delta_0$ into Eqs. (\ref{25}). Then we derive the displacement field from Eqs. (\ref{27})
\begin{equation}
\begin{aligned}
&u_1=S_{11}u_{11}+S_{13}u_{12}+u_{13},
\\&u_3=S_{31}u_{31}+S_{33}u_{32}+u_{33}.
\label{29}
\end{aligned}
\end{equation}
$u_{13}$ is a function of $z$, $u_{33}$ is a function of $x$ and

\begin{eqnarray}
u_{11}=&\frac {\beta^2}{4\pi}\sqrt{\frac{1+\mu '}{1-\mu '}}\mathrm {ln}\left(\frac{z'^2+x^2+xz'\sqrt{2(1-\mu ')}}{z'^2+x^2-xz'\sqrt{2(1-\mu')}}\right)\nonumber\\
&-\frac{\beta^2}{2\pi}\mathrm{arctan}\left(\frac{xz'\sqrt{2(1+\mu')}}{z'^2-x^2}\right)\nonumber\\
&+\frac{\beta^2}2\left(\mathrm H_{z'}(2z'-x)-\mathrm H_{-z'}(x)\right),\nonumber\\
u_{12}=&-\frac{1}{2\pi}\mathrm{arctan}\left(\frac{xz'\sqrt{2(1+\mu')}}{z'^2-x^2}\right)\nonumber\\
&-\frac 1{4\pi}\sqrt{\frac{1+\mu'}{1-\mu'}}\mathrm {ln}\left(\frac{z'^2+x^2+x'z\sqrt{2(1-\mu')}}{z'^2+x^2-xz'\sqrt{2(1-\mu')}}\right)\label{30}\\
&+\frac12(\mathrm H_{z'}(2z'-x)-\mathrm H_{-z'}(x)),\nonumber\\
u_{31}=&-\frac{\beta}{\pi}\sqrt{\frac{1}{2(1-\mu')}}\mathrm{arctan}\left(\frac{\sqrt{1-\mu'^2}z'^2}{x^2+\mu'z'^2}\right),\nonumber\\
u_{32}=&-\frac{\sqrt{2(1+\mu')}}{4\pi\beta}\mathrm{ln}(z'^4+x^4+2z'^2x^2\mu)\nonumber\\
&-\frac{\mu'}{\pi\beta\sqrt{2(1-\mu')}}\mathrm{arctan}\left(\frac{\sqrt{1-\mu'^2}x^2}{z'^2+\mu'x^2}\right),\nonumber
\end{eqnarray}

where $\mathrm H_{z'} (x)$ is defined as $\mathrm H_{z'} (x)=\mathrm H(x-z')$,with $\mathrm H(x)=\frac{1+\mathrm{sgn}(x)}2$ the Heaviside step function. The Heaviside step functions are added in formulae (\ref{30}) to ensure the continuity of displacement field on points $x=z'$ and $x=-z'$.

By substituting Eqs. (\ref{29}) and (\ref{30}) into Eq. (\ref{28}), we get the following differential equation with a very simple form
\begin{equation}
\frac{du_{13} (z)}{dz}+\frac{du_{33} (x)}{dx}=0,
\label{31}
\end{equation}
which has the solution $u_{13}=kz+m, u_{33}=-kx+n$, with $k,\;m$ and $n$ constants. The meaning of $k$ is that the material rotates around $y$-axis with an angle $-\mathrm{arctan}(k)$, and then enlarges it's volume  $(1+k^2)^{\frac32}$ times. $m$ and $n$ represent the rigid body movement. Set $k=m=n=0$, we have
\begin{equation}
\begin{aligned}
&u_1=S_{11}u_{11}+S_{13}u_{12},\\
&u_3=S_{31}u_{31}+S_{33}u_{32}.
\label{32}
\end{aligned}
\end{equation}
Consequently, the displacement field for arbitrary surface force distribution $Q(x)$ can be easily obtained, from Eqs. (\ref{32})
\begin{equation}
\begin{aligned}
&u_1=\int_{-\infty}^{+\infty}(S_{11}u_{11}(\xi,z)+S_{13}u_{12}(\xi,z))Q(x-\xi)d\xi,\\
&u_3=\int_{-\infty}^{+\infty}(S_{31}u_{31}(\xi,z)+S_{33}u_{32}(\xi,z))Q(x-\xi)d\xi.
\label{33}
\end{aligned}
\end{equation}

By using the same method, we can derive the displacement field induced by the shear force distribution $P(x)$ as:
\begin{equation}
\begin{aligned}
&u_1=\int_{-\infty}^{+\infty}(S_{11}u'_{11}(\xi,z)+S_{13}u'_{12}(\xi,z))Q(x-\xi)d\xi,\\
&u_3=\int_{-\infty}^{+\infty}(S_{31}u'_{31}(\xi,z)+S_{33}u'_{32}(\xi,z))Q(x-\xi)d\xi.
\end{aligned}
\end{equation}
where
\begin{equation}
\begin{aligned}
u'_{11}=&-\frac{\beta^2\mu'}{\pi\sqrt{2(1-\mu')}}\mathrm{arctan}\left(\frac{\sqrt{1-\mu'^2}z'^2}{x^2+\mu'z'^2}\right)\\
&-\frac{\beta^2\sqrt{2(1+\mu')}}{4\pi}\mathrm{ln}(z'^4+x^4+2z'^2x^2\mu'),\\
u'_{12}=&-\frac{u_{31}}\beta,\;\;  u'_{31}=-\frac{u_{11}}\beta, \;\; u'_{32}=-\frac{u_{12}}\beta.\\
\end{aligned}
\end{equation}

We now consider a simple case when $\theta=0$ and the semi-infinite cubic crystal is subjected to an evenly distributed normal force on the surface, $Q=1$. Obviously, the displacement field along $z$-axis is linear with $z$: $u_3 (x,z)=kz$ (solution 1), where $k$ is a constant merely related to elastic moduli. But on the other hand, via the formulae (\ref{33}), one arrives at
\begin{equation}
u_3(x,z)=\int_{-\infty}^{+\infty}(S_{31}u_{31}(\xi,z)+S_{33}u_{32}(\xi,z))d\xi,
\label{34}
\end{equation}
(solution 2). We find that solution 1 and solution 2 are not the same; moreover, the integral form of solution 2 is divergent. In fact, the difference between those two solutions originates from the choice of the fixed plane: solution 1 is obtained under the assumption that the plane $z=0$ is fixed, while solution 2 is gotten with the plane $z=+\infty$ fixed. According to the theory of elasticity, such difference (even though infinite) can be seen as a constant. To eliminate this special constant, we calculate the finite part of the divergent integral (\ref{34}) by using the method of functional regularization of general function which regards the order of differential and integral as exchangeable \cite{25}. We first calculate the partial derivative of solution 2 with respect to $z$, and then, integrate the obtained partial derivative with respect to $z$. The result, $u_3 (x,z)=-\frac{z}{C_{11}-C_{12}}$, has the same form as solution 1. Thus, from a physics point of view, the mathematical difficulty is just due to the choice of reference system, and it can be solved by translating the reference system along $z$-axis with an infinite distance. Mathematically, the method is related to the calculation of the finite part of the divergent integral.

\section{Analytical solution of surface-induced displacement field for skyrmion phase}

For $Q=F\mathrm{cos}(ax)$, the displacement field is expected to be periodic. The general formulae (\ref{33}) give
\begin{equation}
\begin{aligned}
&u_1=F\int_{-\infty}^{+\infty}(S_{11}u_{11}(\xi,z)+S_{13}u_{12}(\xi,z))\mathrm{cos}(a(x-\xi))d\xi,
\\&u_3=F\int_{-\infty}^{+\infty}(S_{31}u_{31}(\xi,z)+S_{33}u_{32}(\xi,z))\mathrm{cos}(a(x-\xi))d\xi,
\label{35}
\end{aligned}
\end{equation}
which are divergent. The derivatives of functions (\ref{35}) with respect to $x$ are 
\begin{equation}
\begin{aligned}
&\frac{\mathrm{\partial} u_1(x,z)}{\mathrm{\partial} x}=Ff_{1,\theta} (z)\mathrm{cos}(ax),
\\&\frac{\partial u_3(x,z)}{\partial x}=-Ff_{2,\theta} (z)\mathrm{sin}(ax),
\end{aligned}
\label{36}
\end{equation}
where $f_{1,\theta} (z)$ and $f_{2,\theta} (z)$ are expressed in Eqs. (\ref{37}) with $a_1=\sqrt{\frac{1-\mu'}2}\beta a,\;a_2=\sqrt{\frac{1+\mu'}2}\beta a$. By integrating functions (\ref{36}) with respect to $x$, the displacement field is derived as
\begin{equation}
\begin{aligned}
&u_1(x,z)=\frac Faf_{1,\theta} (z)\mathrm{sin}(ax),
\\&u_3(x,z)=\frac Faf_{2,\theta} (z)\mathrm{cos}(ax),
\label{38}
\end{aligned}
\end{equation}
which is composed of two parts: one is the harmonic term having the same period as the force distribution, the other is the $z$-related term having an exponential factor $\mathrm e^{a_2z}$. $a_2=\sqrt{\frac{1+\mu'}2}\beta a$ is positive; therefore, the displacement decreases rapidly with decreasing $z$. For the region far away from the surface, i.e., the distance from the boundary greater than several times of wavelength of the harmonic force distribution, the displacement is null. Thus, the elastic field derived in Ref. \cite{20} is suitable for bulk materials.

\begin{widetext}
\begin{equation}
\begin{aligned}
&f_{1,\theta} (z)=\left(S_{11}\beta^2 \left(\mathrm{cos}(a_1 z)+\sqrt{\frac{1+\mu'}{1-\mu'}}\mathrm{sin}(a_1 z) \right)+S_{13} \left(\mathrm{cos}(a_1 z)-\sqrt{\frac{1+\mu'}{1-\mu'}} \mathrm{sin}(a_1 z)\right)\right) \mathrm e^{a_2z},
\\&f_{2,\theta} (z)=\sqrt{2(1+\mu')}\left(S_{31}\beta\left(\frac1{\sqrt{1-\mu'^2}} \mathrm{sin}(a_1 z)\right)+\frac{S_{33}}{\beta} \left(\mathrm{cos}(a_1 z)-\frac {\mu'}{\sqrt{1-\mu'^2}}\mathrm{sin}(a_1 z) \right)\right)\mathrm e^{a_2z}.
\label{37}
\end{aligned}
\end{equation}
\begin{equation}
\begin{aligned}
&f_{3,\theta} (z)=\sqrt{2(1+\mu')}\left(S_{11}\beta^2 \left(\mathrm{cos}(a_1 z)+\frac {\mu'}{\sqrt{1-\mu'^2}}\mathrm{sin}(a_1 z) \right)-S_{13}\left(\frac1{\sqrt{1-\mu'^2}} \mathrm{sin}(a_1 z)\right)\right)\mathrm e^{a_2z},\\
&f_{4,\theta} (z)=\left(S_{31}\beta\left(\mathrm{cos}(a_1 z)+\sqrt{\frac{1+\mu'}{1-\mu'}}\mathrm{sin}(a_1 z) \right)+\frac{S_{33}}{\beta} \left(\mathrm{cos}(a_1 z)-\sqrt{\frac{1+\mu'}{1-\mu'}} \mathrm{sin}(a_1 z)\right)\right) \mathrm e^{a_2z}.\\
\label{371}
\end{aligned}
\end{equation}
\begin{equation}
\begin{aligned}
u_1=&-\sum_{i,j,k=1}^3\left(\frac{f_1^{ij}(\bm e_1\cdot \bm q_{ij})}{|\bm q_{ij}|^2}+\frac{f_3^{ij}{\bm \left(\bm e_k\cdot\bm q_{ij}\right)\left(\bm e_1\cdot\bm q_{ij}\right)}}{|\bm q_{ij}|^3}\mathrm i+\frac{f_5^{ij}((\bm q_{ij}\times \bm e_k)\cdot \bm e_3)((\bm q_{ij}\times \bm e_1)\cdot \bm e_3)}{|\bm q_{ij}|^3}\mathrm i\right)\sigma_{k3}^{Sij}\mathrm {sin}(\bm r\cdot \bm q_{ij}),\\
u_2=&-\sum_{i,j,k=1}^3\left(\frac{f_1^{ij}(\bm e_2\cdot \bm q_{ij})}{|\bm q_{ij}|^2}+\frac{f_3^{ij}{\bm \left(\bm e_k\cdot\bm q_{ij}\right)\left(\bm e_2\cdot\bm q_{ij}\right)}}{|\bm q_{ij}|^3}\mathrm i+\frac{f_5^{ij}((\bm q_{ij}\times \bm e_k)\cdot \bm e_3)((\bm q_{ij}\times \bm e_2)\cdot \bm e_3)}{|\bm q_{ij}|^3}\mathrm i\right)\sigma_{k3}^{Sij}\mathrm {sin}(\bm r\cdot \bm q_{ij}),\\
u_3=&-\sum_{i,j,k=1}^3\left(\frac{f_2^{ij}(\bm e_k\cdot\bm e_3)}{|\bm q_{ij}|}+\frac{f_4^{ij}(\bm e_k\cdot\bm q_{ij})}{|\bm q_{ij}|}\mathrm i\right)\sigma_{k3}^{Sij}\mathrm{cos}(\bm r\cdot\bm q_{ij}).
\label{372}
\end{aligned}
\end{equation}
\end{widetext}

Similarly, for $P=F\mathrm{sin}(ax)$,  we can derive the displacement field as:
\begin{equation}
\begin{aligned}
&u_1(x,z)=\frac Faf_{3,\theta} (z)\mathrm{sin}(ax),
\\&u_3(x,z)=\frac Faf_{4,\theta} (z)\mathrm{cos}(ax),
\end{aligned}
\end{equation}
with $f_{3,\theta} (z)$ and $f_{4,\theta} (z)$ expressed in Eqs. (\ref{371}).

We have solved the 2D displacement field for surface forces with distribution along $x$-axis and with direction along $x$-axis ($Q$) and $z$-axis ($P$). As to the 3D case, it is induced by another kind of surface force, which we denote as $R$, with distribution along $x$-axis and with direction along $y$-axis. For $R=F\mathrm {sin}(ax)$, we give directly the displacement field as
\begin{equation}
u_1=u_3=0,~~u_2(x,z)=\frac Faf_{5,\theta}(z)\mathrm{sin}(ax).
\end{equation}
Here, $f_{5,\theta}(z)=S_{55}\mathrm e^{az}$.

	The $X$, $Y$ and $Z$-direction forces $F_1=-\sigma_{13}$, $F_2=-\sigma_{23}$ and $F_3=-\sigma_{33}$ are composed of nine $\bm q_{ij}$ (see Eq. (\ref{A3})) structures. For each $\bm q_{ij}$ structure of $F_1$ or $F_2$, the components in the direction of and perpendicular to $\bm q_{ij}$ are $P$-type and $R$-type forces, respectively. For $\bm q_{ij}$  structures of $F_3$, they are $Q$-type forces. Solving the displacement field for each $\bm q_{ij}$  structure in corresponding O-xyz system and projecting it onto the $X$, $Y$ and $Z$-axes, we can finally get the surface-induced displacement field in O-XYZ system as in Eqs. (\ref{372}), where $f_k^{ij}\;(k=1,..., 5)$ takes the value of $f_{k,\theta}(0)$ for $\theta=\mathrm {arccos }\left(\frac{\bm e_1\cdot\bm q_{ij}}{|\bm q_{ij}|}\right)$, $\bm e_1$, $\bm e_2$ and $\bm e_3$ are the unit vectors along $X$, $Y$ and $Z$-axis respectively. 

For helimagnet MnSi, the related parameters are: $C_{11}=2.83\times 10^{11}$ Pa, $C_{12}=0.641\times 10^{11}$ Pa, $C_{44}=1.179\times 10^{11}$ Pa \cite{23}, $K=-2\times 10^7 $ JA$^{-2}$m$^{-1}$, $L_1=-0.7\times 10^6$ JA$^{-2}$m$^{-1}$, $L_2=0.6\times 10^6$ JA$^{-2}$m$^{-1}$, $L_3=1.646\times 10^6$ JA$^{-2}$m$^{-1}$, $L_{O1}=1.147\times 10^{-4}$ JA$^{-2}$m$^{-2}$, $L_{O2}=-0.537\times 10^{-4}$ JA$^{-2}$m$^{-2}$, $L_{O3}=-0.537\times 10^{-4}$ JA$^{-2}$m$^{-2}$, $L_{O4}=L_{O5}=L_{O6}=0$ \cite {22}, and $q=\frac{|b|}{2A}=4.5\times 10^{8}$ m$^{-1}$ \cite{1,53}. According to the analytical expressions of surface-induced displacement field in Eqs. (\ref{372}) and skyrmion-induced displacement field in Ref. \cite {20}, the contour maps of the displacement components at 4 K and 0.1 T are plotted in FIG. 2. At the center and the six vertexes of a skyrmion lattice, there appear the peaks, for which the $X$ and $Y$-components of the total displacement, $u_1^t$ and $u_2^t$, are zero; while the $Z$-component, $u_3^t$, takes a maximum value. At the right-hand part and upper part of a peak, we have $u_1^t>0$ and $u_2^t>0$, respectively; this indicates the tendency of expansion of the peaks. $u_1^t$ and $u_2^t$ are a little deformed, to explain this, the skyrmion-induced, normal-force-induced and shear-force-induced $X$-direction displacements $u_1^{sky}$,$u_1^{nor}$ and $u_1^{she}$ are plotted as (d), (e) and (f) in FIG. 2, respectively. $u_1^{sky}$ and $u_1^{nor}$ share the same pattern with zero-value contour lines along $Y$-axis; while $u_1^{she}$ shows different behavior with zero-value contour lines along $X$-axis. It is the shear force who deforms $u_1^t$. $u_1^t$ is larger than $u_1^{sky}$. The skyrmion-induced elastic stresses tend to decrease the total displacement, while at the surface they are released. Therefore, the total displacement increases at the surface.

\begin{figure}
\centering
\includegraphics[scale=0.28]{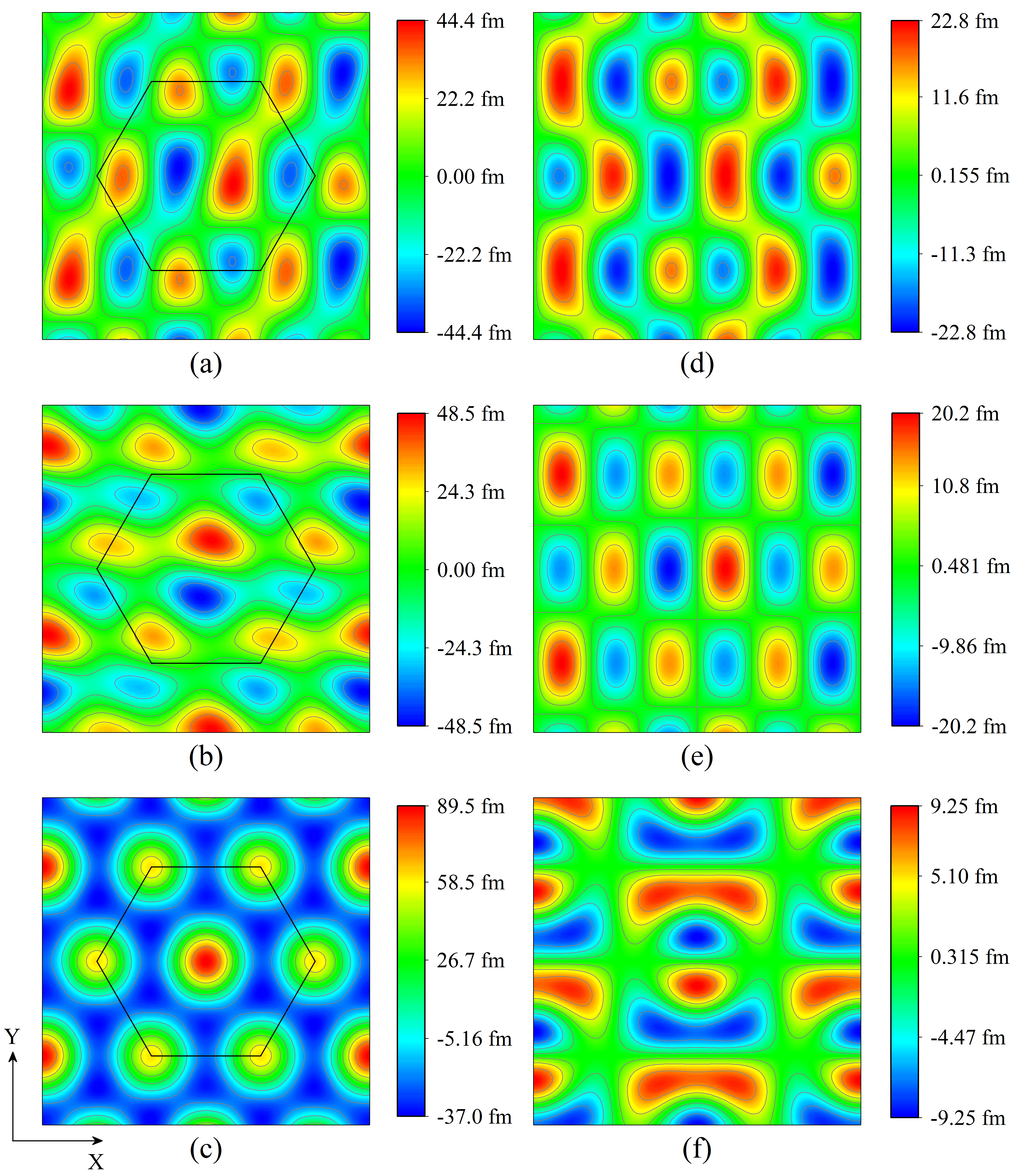}
\caption{Contour plots of displacement components at 4 K and 0.1 T. The region enclosed by the hexagon represents a skyrmion lattice. (a), (b) and (c) stand for the total displacements along $X$, $Y$ and $Z$-direction respectively; (d), (e) and (f) show the skyrmion-induced, normal-force-induced and shear-force-induced $X$-direction displacements, respectively.}
\end{figure}

\section{Discussion}
\subsection{Tunability of surface configuration by bias magnetic field}

We plot the surface configuration of skyrmions at 4 K and under different applied magnetic field $B$. FIG. 3 (a-d) represent the total normal surface displacement field at 0.1 T, 0.2 T, 0.3 T and 0.4 T respectively. At 0.2 T, the surface is characterized by peaks (arranged periodically like the triangular skyrmion lattices) with almost the same height . For $B>0.2$ T, the center peak is higher than the six adjacent peaks, while for $B<0.2$T, the reverse is the case, indicating that the heights of these two types of peaks compete with each other.

\begin{figure}
\centering
\includegraphics[scale=0.135]{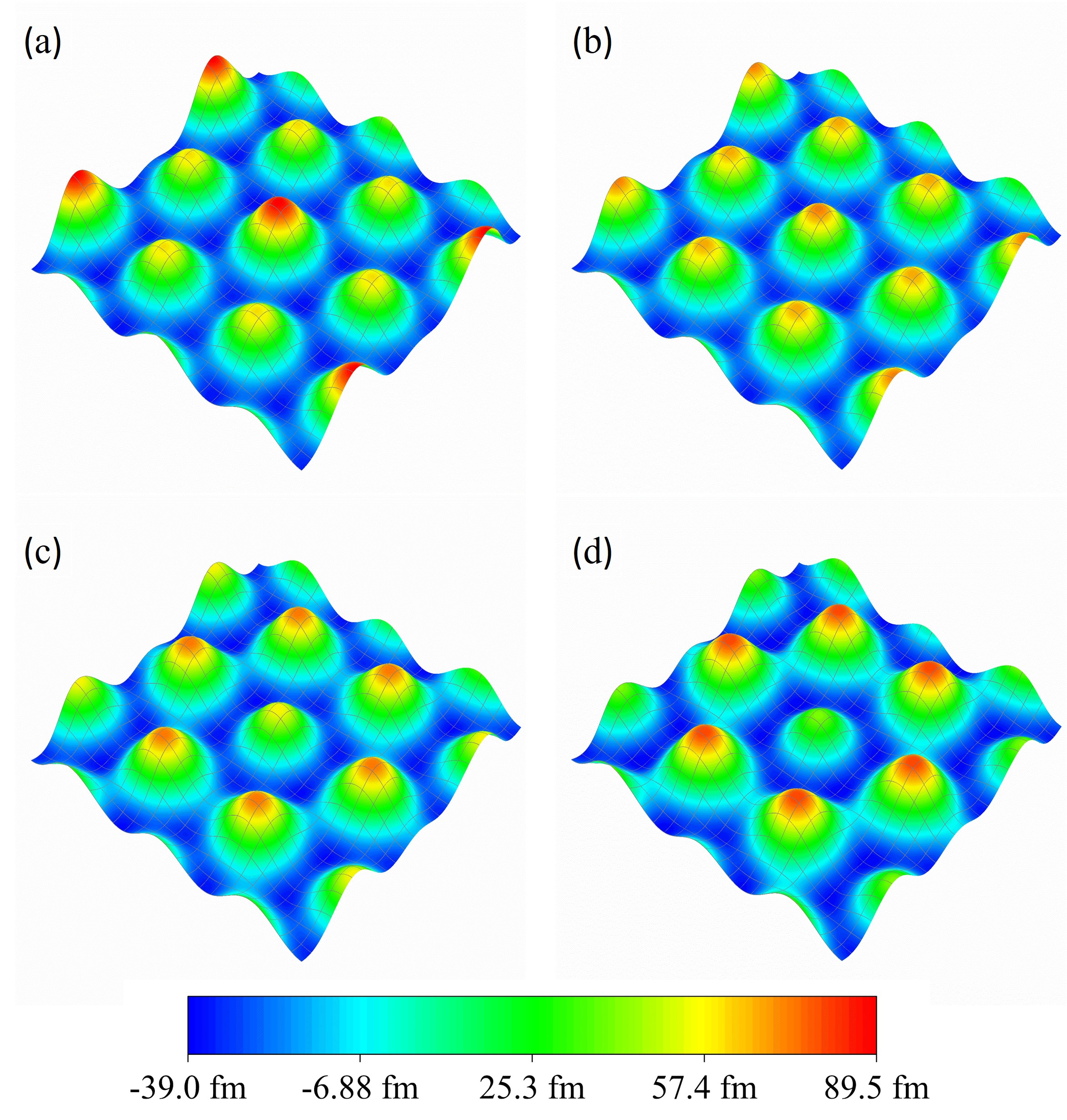}
\caption{Surface configuration of $u_3$ for MnSi in skyrmion phase at temperature 4K and magnetic field (a) 0.1 T, (b) 0.2 T, (c) 0.3 T and (d) 0.4 T. The size of (a)-(d) is $\frac{2\pi}{q}\times \frac{2\pi}{q}$.}
\end{figure}

To explain the competing behavior of these two patterns of peaks, we explore separately the two dominant parts of the displacement: the $\sigma_{33}^{S1}$-induced normal displacement $u_3^{s1}$ and the $\sigma_{33}^{S3}$-induced normal displacement $u_3^{s3}$. FIG. 4 shows the surface displacement $u_3^{s1}$ at 0.1 T, 0.2 T, 0.3 T and 0.4 T. It can be seen that $u_3^{s1}$ goes through the same ``configurational reversal'' as $\sigma_{33}^{S1}$ when the external magnetic field increases. At 0.1 T, there are periodically arranged peaks on the surface. With the augmentation of the magnetic field, the height of the peaks decreases, then at about 0.2 T, when tan$(\varphi)\approx 2.35$, the peaks vanishes, and the surface described by $u_3^{s1}$ becomes almost flat. For $B>0.2$ T, on the surface, there appears the valleys, the depth of which increases when the magnetic field augments. The ``configurational reversal'' can be explained through the relation between $u_3^{s1}$ and $\sigma_{33}^{S1}$ revealed by Eqs. (\ref{372}). As for $u_3^{s3}$, Eqs. (\ref{372}) and the invariability of $\sigma_{33}^{S3}$ imply that $u_3^{s3}$ keeps almost unchanged when magnetic field changes. It is the reversal feature of $u_3^{s1}$ and the invariability of $u_3^{s3}$ that decide the competing behavior of two patterns of peaks.

\begin{figure}
\centering
\includegraphics[scale=0.135]{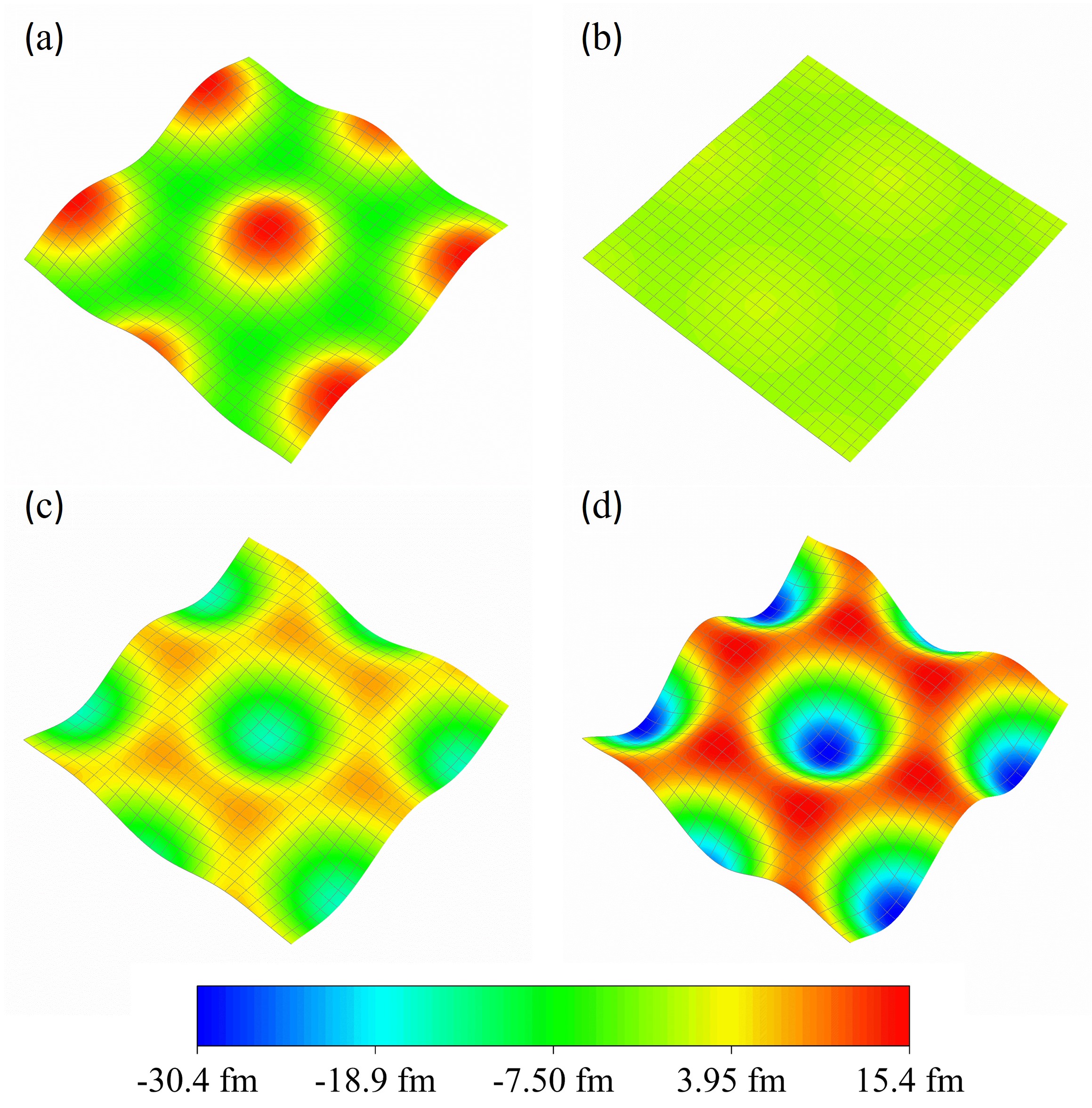}
\caption{Reversible surface configuration of $u_3^{s1}$ for MnSi in skyrmion phase at temperature 4 K and magnetic field (a) 0.1 T, (b) 0.2 T, (c) 0.3 T and (d) 0.4 T.}
\end{figure}

According to Eqs. (\ref{372}), the displacement field $u_3$ can be divided into three triple-Q structures: $u_3^{Q1} $, $u_3^{Q2}$ and $u_3^{Q3}$, corresponding to $\bm q_{1i}$, $\bm q_{2i}$ and $\bm q_{3i} \;(i=1,2,3)$, respectively. 
To explore the periodicity of $u_3$, we plot the simplest repeating unit of surface displacement $u_3^{Q1}$, $u_3^{Q2}$, $u_3^{Q3}$ and $u_3$ at 4 K and 0.1 T in FIG. 5. We can see that $u_3^{Q1}$ and $u_3$ share the same periodicity. The primitive vectors for the hexagonal lattices of $u_3^{Qi}$ are $\bm a_{i1}$ and $\bm a_{i2}$, satisfying $\bm a_{ij}\cdot\bm q_{ik}=2\pi \delta_{jk}$ where  $i=1,2,3$ and $j,k=1,2$, $\delta_{ij}$ is the Kronecker delta. We can demonstrate that $\bm a_{11}=2\bm a_{21}=\bm a_{31}+2\bm a_{32}$ and $\bm a_{12}=2\bm a_{22}=-\bm a_{31}+\bm a_{32}$. Thus, for arbitrary integers $n_1$ and $n_2$, we have $u_3^{Q2} (\bm r+n_1 \bm a_{11}+n_2 \bm a_{12} )=u_3^{Q3} (\bm r+2(n_1+n_2 ) \bm a_{21}+2(n_1+n_2 ) \bm a_{22} )=u_3^{Q2} (\bm r)$ and $u_3^{Q3} (\bm r+n_1 \bm a_{11}+n_2 \bm a_{12} )=u_3^{Q3} (\bm r+(n_1-n_2 ) \bm a_{31}+(2n_1+n_2 ) \bm a_{32} )=u_3^{Q3} (\bm r)$. Consequently, $u_3$ has the same period as $u_3^{Q1}$ and the skyrmion lattices. By using the relations between $a_{ij}$, $u_1$ and $u_2 $ can also be demonstrated to share the same periodicity as the skyrmion lattices. 

\begin{figure}
\centering
\includegraphics[scale=0.11]{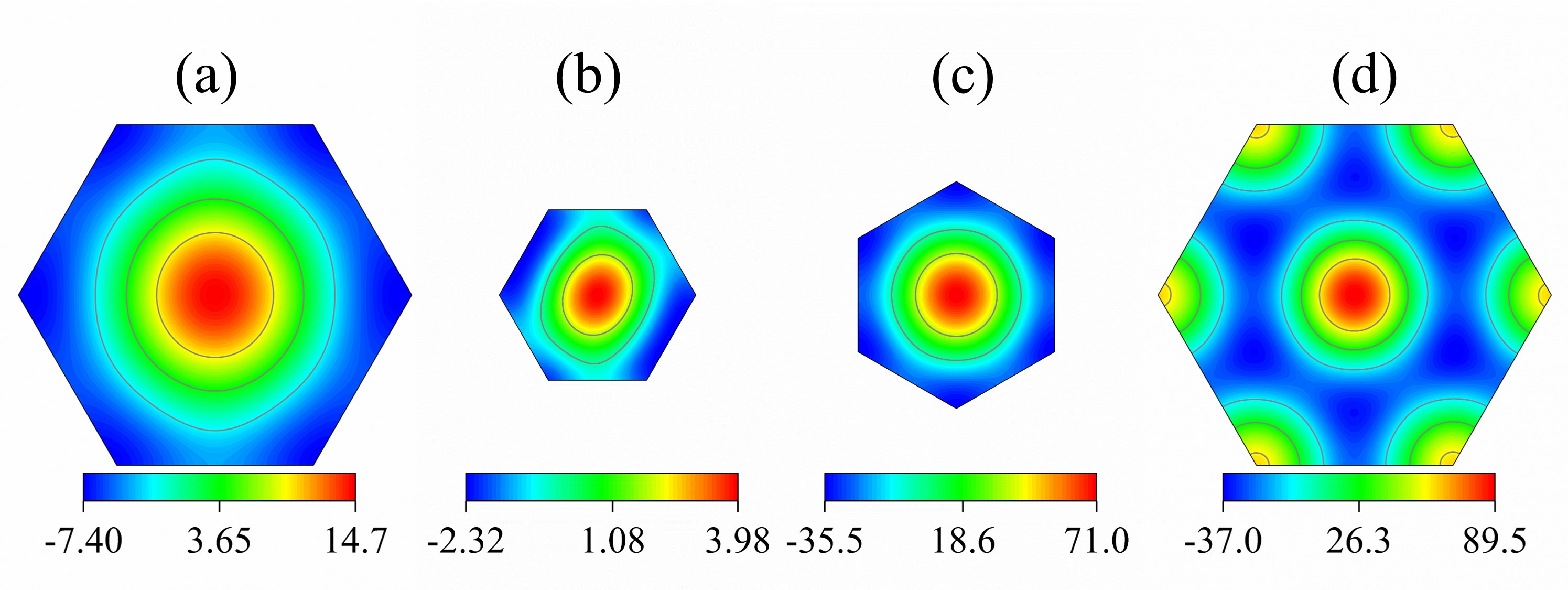}
\caption{The simplest repeating unit of surface displacement (the unit is fm). (a) $u_3^{Q1}$, (b) $u_3^{Q2}$, (c) $u_3^{Q3}$ and (d)  $u_3$ for MnSi in skyrmion phase at 4 K and 0.1 T.}
\end{figure}

\subsection{Possible effects of electric current and mechanical load on the surface configuration}
It is known that skyrmions behave like moving particles with stable topological structures when exposed to various kinds of external fields including electric current \cite{10,26} and temperature gradient \cite{27}. A further concern is how will the surface configuration change with the motion of skyrmions. For moving skyrmions at speed $\bm v$, the magnetization can be described by introducing a translation transformation: $\bm r\rightarrow\bm r-\bm vt$. Thus, we have $\bm M=\bm M(\bm r-\bm vt)$, where $\bm M(\bm r)$ is expressed as Eq. (\ref{5}). Correspondingly, the solution of $u_i \;(i=1,2,3)$ obtained in Eqs. (\ref{372}) is changed by replacing $\bm r$ with $\bm r-\bm vt$, i.e. $u_i=u_i(\bm r-\bm vt)$. Thus, the displacement field moves together with skyrmions.

When anisotropic mechanical loads are applied to helimagnets, skyrmion lattices are found to undergo emergent elastic deformation independent of the deformation of the underlying atomic lattices \cite{19}. It is shown in Ref. \cite{28} that the deformed skyrmions have a triple-Q structure characterized by $\bm q_1,\; \bm q_2$ and $\bm q_3$ satisfying $|\bm q_1|\neq |\bm q_2|\neq |\bm q_3|$ and  $\bm q_1+\bm q_2+\bm q_3=\bm0$. For a general analysis, we can see that the periodic eigenstrains obtained from Eq. (\ref{4}) is still composed of three triple-Q structures. The periodic stress field, linearly related to the incompatible part of eigenstrains, obviously shares the same periodicity with the eigenstrains. From Eqs. (\ref{372}), we can see that for arbitrarily deformed skyrmion lattices, $u_3^{Q1}$ and $u_3^{Q2}$ has the same periodicity with the deformed skyrmions, while $u_3^{Q3}$ is a triple-Q structure with the three ``Q''s: $\bm q_1-\bm q_2$, $\bm q_1-\bm q_3$ and $\bm q_2-\bm q_3$. Following the proof given in part $A$ of this section, we can easily show that $u_3^{Q3}$ and $u_3$ share the same periodicity, because $\bm q_1+\bm q_2+\bm q_3=\bm0$ is the only necessary condition which is still valid for any deformed skyrmion lattices. Therefore, the surface displacement field deforms together with the skyrmion lattices.

We have proved qualitatively that the surface displacement field moves together, and deforms together with the skyrmion lattices. Therefore, the various kinds of approaches discovered to affect the skyrmion lattices will also be effective in controlling the surface displacement field.

\subsection{Generality and possible technological interest}
Apart from two-dimensional DM-induced Bloch-type magnetic skyrmion lattices in helimagnets, skyrmions can exist in many other forms: three-dimensional skyrmions, such as hourglass-shaped skyrmions \cite{29} and bobber-shaped skyrmions \cite{30}; atomic-scale skyrmions induced by four-spin interaction \cite{31}, skyrmion bubbles induced by dipole-dipole interaction \cite{55,57}and stabilized by uniaxial anisotropy \cite{54,56}; N\'eel-type skyrmions \cite{32}; isolated skyrmion and skyrmion glass structure \cite{14}. Since magnetoelastic coupling is intrinsic for any ferromagnets, these skyrmions forms are all accompanied by a surface displacement field. Thus, the surface configuration is an additional particle-like property of any magnetic skyrmions.

The maximum displacement perpendicular to the surface is of the order of magnitude of $10^{-13}$ m for MnSi. Such a small displacement is difficult to detect. But as shown in formulae (\ref{7})-(\ref{15}), and (\ref{372}), the displacement is related to the magnetoelastic coefficients, and the size of skyrmion lattices. To get a greater displacement, one should pay attention to materials hosting skyrmions with bigger size and having stronger magnetoelastic coupling, for instance, FeGe. Even though the magnetoelastic coefficients are not available due to the technical difficulties in fabricating large FeGe single crystals \cite{33}, one can expect to observe larger displacement field for FeGe than for MnSi. The skyrmion lattice parameter for FeGe is about 70 nm \cite{34}, four times larger than that for MnSi. Moreover, the experiment carried out by K. Shibata $et\; al.$ \cite{19}, in which anisotropic strain as small as $0.3\%$ induced distortions of skyrmion lattices by $20\%$, implies large magnetoelastic coupling in FeGe.

\section{Conclusion}
We have obtained the analytical solution of displacement field at the surface of cubic helimagnets in skyrmion phase. For MnSi, The normal displacement field is dominated by two triple-Q structures $u_3^{s1}$ and $u_3^{s3}$ . $u_3^{s3}$  is characterized by periodically arranged peaks having invariant height when applied magnetic field changes and $u_3^{s1}$ , undergoing a ``configurational reversal'' when the magnetic field increases from 0.1 T to 0.4 T, distinguishes these peaks into two patterns which compete with each other. The surface configuration enriches the meaning of particle-like nature of magnetic skyrmions, it moves and deforms with the skyrmions lattices and can be therefore controlled by applied field, such as magnetic field, current etc.

\begin{acknowledgments}
The work was supported by the NSFC (National Natural Science Foundation of China) through the fund 11302267.
\end{acknowledgments}

$Author\;contributions$:
Y. Hu conceived the idea. X. Wan finished the analytical deduction. X. Wan, Y. Hu and B. Wang discussed the results for revision. X. Wan, Y. Hu and B. Wang co-wrote the manuscript.

\begin{appendix}

\section {Analytical solution of the skyrmion-induced stress field}
For a bulk cubic crystal free from body forces and surface constraints, we obtain the expressions of eigenstrains $\varepsilon_{ij}^*=\varepsilon_{ij}^*(\bm M) \; (i,j=1,2,3)$ by solving the equations $\sigma_{IJ} (\varepsilon_{ij}^*,\bm M)=0  \; (I,J,i,j=1,2,3) $, where $\sigma_{IJ}$, a function of $\varepsilon_{ij} (i,j=1,2,3)$ and $\bm M$, is obtained by $\sigma_{IJ} (\varepsilon_{ij},\bm M)=\frac{\partial w}{\partial {\varepsilon_{IJ}}}   \; (\text{for } I=J) \text { and } \sigma_{IJ} (\varepsilon_{ij},\bm M)=\frac{\partial w}{\partial{\gamma_{IJ}}} \; (\text{for } I\neq J)$.
\begin{widetext}
\begin{equation}
\begin{aligned}
&\varepsilon_{11}^*=K^*M^2-L_1^*M_1^2-L_2^*M_3^2+L_{O1}^*(M_3 M_{1,2}-M_2 M_{1,3} )+L_{O2}^* (M_3 M_{2,1}-M_2 M_{3,1 })+L_{O3}^* M_1 (M_{2,3}-M_{3,2}),
\\&\varepsilon_{22}^*=K^*M^2-L_1^*M_2^2-L_2^*M_1^2+L_{O1}^*(M_1 M_{2,3}-M_3 M_{2,1} )+L_{O2}^* (M_1 M_{3,2}-M_3 M_{1,2 })+L_{O3}^* M_2 (M_{3,1}-M_{1,3}),
\\&\varepsilon_{33}^*=K^* M^2-L_1^* M_3^2-L_2^* M_2^2+L_{O1}^* (M_2 M_{3,1}-M_1 M_{3,2} )+L_{O2}^* (M_2 M_{1,3}-M_1 M_{2,3} )+L_{O3}^* M_3 (M_{1,2}-M_{2,1} ),
\\&\gamma_{2,3}^*=\frac 1{C_{44}M^2_s}[-L_3 M_2 M_3+L_{O6} M_1 (M_{2,2}-M_{3,3} )+M_2 (L_{O4} M_{1,2}+L_{O5} M_{2,1} )-M_3 (L_{O4} M_{1,3}+L_{O5} M_{3,1} )],
\\&\gamma_{1,3}^*=\frac 1{C_{44}M^2_s}[-L_3 M_1 M_3+L_{O6} M_2 (M_{3,3}-M_{1,1} )+M_3 (L_{O4} M_{2,3}+L_{O5} M_{3,2} )-M_1 (L_{O4} M_{2,1}+L_{O5} M_{1,2} )],
\\&\gamma_{1,2}^*=\frac 1{C_{44}M^2_s}[-L_3 M_1 M_2+L_{O6} M_3 (M_{1,1}-M_{2,2} )+M_1 (L_{O4} M_{3,1}+L_{O5} M_{1,3} )-M_2 (L_{O4} M_{3,2}+L_{O5} M_{2,3} )],
\end{aligned}
\label{4}
\end{equation}
\end{widetext}
with $K^*=\frac{-C_{11}K+C_{12}(K+L_1+L_2)}{(C_{11}-C_{12})(C_{11}+2C_{12})M_s^2}$, $L_1^*=\frac{L_1}{(C_{11}+2C_{12})M_s^2}$, $L_2^*=\frac{L_2}{(C_{11}+2C_{12})M_s^2}$, $L_{O1}^*=\frac{-C_{11}L_{O1}+C_{12}(K+L_{O1}+L_{O2})}{(C_{11}-C_{12})(C_{11}+2C_{12})M_s^2}$, $L_{O2}^*=\frac{-C_{11}L_{O2}+C_{12}(K+L_{O1}+L_{O2})}{(C_{11}-C_{12})(C_{11}+2C_{12})M_s^2}$ and $L_{O3}^*=\frac{-C_{11}L_{O3}+C_{12}(K+L_{O1}+L_{O2})}{(C_{11}-C_{12})(C_{11}+2C_{12})M_s^2}$.

By substituting Hooke's law, describing the linear relation between stresses $\sigma_{ij}$ and elastic strains $e_{ij}$, which is the difference between total strains $\varepsilon_{ij}$ and eigenstrains $\varepsilon_{ij}^*$ , and geometrical equations $\varepsilon_{ij}=\frac{u_{i,j}+u_{j,i}}2$ into the equilibrium equations, we obtain three partial differential equations about the displacements $u_i$

\begin{equation}
\begin{aligned}
&C_{11}u_{i,ii}+C_{44}(u_{i,jj}+u_{i,kk})+(C_{12}+C_{44})(u_{j,ij}+u_{k,ik})\\&=C_{11}\varepsilon_{ii,i}^*+C_{12}(\varepsilon_{jj,i}^*+\varepsilon_{kk,i}^*)+C_{44}(\gamma_{ij,j}^*+\gamma_{1k,k}^*),\\
\end{aligned}
\end{equation}
where $i,j,k=1,2,3$ and $i\neq j\neq k$.

 $\varepsilon_{ij}^*=\varepsilon_{ij}^* (\bm M)$ and $\gamma_{ij}^*=\gamma_{ij}^* (\bm M)$ are quadratic functions of $\bm M$ \cite{20}. By substituting the triple-Q periodic form of $\bm M$ into the obtained eigenstrains, we can find that eigenstrains have a multi-Q structure with nine wavevectors $\bm q_{ij}\; (i,j=1,2,3)$ defined as:
\begin{equation}
\left[\bm q_{ij}\right]= \begin{bmatrix} \bm q_1& \bm q_2&\bm q_3\\2\bm q_1&2\bm q_2&2\bm q_3\\\bm q_1-\bm q_2&\bm q_1-\bm q_3&\bm q_2-\bm q_3\end{bmatrix}.
\label{A3}
\end{equation}
This multi-Q structure can be seen as the superposition of three triple-Q structures with different magnitudes $q$, $2q$ and $\sqrt 3 q$. Combining the geometrical equations, eigenstrains and Hooke's law, we then derive the triple-Q structure stresses as :

\begin{equation}
\begin{aligned}
&\sigma_{ij}^{Sk}=Re\left[ \sum_{l=1}^3\sigma_{ij}^{Skl}\mathrm e^{\mathrm i \bm q_{kl}\cdot\bm r}\right] \;(i,j,k=1,2,3)
\label{6}
\end{aligned}
\end{equation}
where
\begin{equation}
\begin{bmatrix} \sigma_{13}^{S11}\\ \sigma_{13}^{S12}\\ \sigma_{13}^{S13} \end{bmatrix}=\frac {\mathrm iL_3M^2} {12M_s^2} \mathrm{sin}(\varphi)(6\mathrm{cos}(\varphi)-\sqrt3 \mathrm{sin}(\varphi))\begin{bmatrix} 0\\1\\-1\end{bmatrix},
\label{7}
\end{equation}
\begin{equation}
\begin{bmatrix} \sigma_{13}^{S21}\\ \sigma_{13}^{S22}\\ \sigma_{13}^{S23} \end{bmatrix}=-\frac {\mathrm i L_3M^2} {4\sqrt3M_s^2} \mathrm{sin}^2(\varphi)\begin{bmatrix} 0\\1\\-1\end{bmatrix},
\label{8}
\end{equation}
\begin{equation}
\begin{bmatrix} \sigma_{13}^{S31}\\ \sigma_{13}^{S32}\\ \sigma_{13}^{S33} \end{bmatrix}=\frac {\mathrm i L_3M^2} {4\sqrt3M_s^2}\mathrm{sin}^2(\varphi)\begin{bmatrix} 1\\-1\\-2\end{bmatrix},
\label{9}
\end{equation}
\begin{equation}
\begin{bmatrix} \sigma_{23}^{S11}\\ \sigma_{23}^{S12}\\ \sigma_{23}^{S13} \end{bmatrix}=\frac {\mathrm iL_3M^2} {12M_s^2}  \mathrm{sin}(\varphi)(2\sqrt3\mathrm{\mathrm{cos}}(\varphi)-\mathrm{sin}(\varphi))\begin{bmatrix} 2\\-1\\-1\end{bmatrix},
\label{10}
\end{equation}
\begin{equation}
\begin{bmatrix} \sigma_{23}^{S21}\\ \sigma_{23}^{S22}\\ \sigma_{23}^{S23} \end{bmatrix}=-\frac {\mathrm iL_3M^2} {12M_s^2}\mathrm{sin}^2(\varphi)\begin{bmatrix} 2\\-1\\-1\end{bmatrix},
\label{11}
\end{equation}
\begin{equation}
\begin{bmatrix} \sigma_{23}^{S31}\\ \sigma_{23}^{S32}\\ \sigma_{23}^{S33} \end{bmatrix}=\frac {\mathrm iL_3M^2} {4M_s^2}\mathrm{sin}^2(\varphi)\begin{bmatrix} 1\\1\\0\end{bmatrix},
\label{12}
\end{equation}
\begin{eqnarray}
&\sigma_{33}^{S11}=-\frac{\mathrm{sin}(\varphi)M^2}{12M_s^2C_{11}} \{ 4\sqrt3\mathrm{cos}(\varphi)[-C_{12}(2K+qL_{O2}\nonumber\\
&+C_{11} (2K+2L_1+qL_{O3})]+\mathrm{sin}(\varphi)[C_{11}(-6K\nonumber\\
&-4L_1+L_2-2qL_{O1}-4qL_{O3})+C_{12}(6K+3L_1\nonumber\\
&-2L_2+3qL_{O1}+3qL_{O3} )]\},\nonumber\\
&\sigma_{33}^{S12}=\sigma_{33}^{S13}=\frac {\mathrm{sin}(\varphi)M^2}{6C_kM_s^2}\{-2\sqrt3 \mathrm{cos}(\varphi)[3C_{11}^2 (2K\nonumber\\
&+2L_1+qL_{O3} )-C_{11} (3C_{12} (4K-q(L_{O1}+L_{O2} ))\nonumber\\
&-10C_{44} (2K+2L_1+qL_{O3} ))+C_{12} (3C_{12} (2K\nonumber\\
&-2L_1+q(L_{O1}+L_{O2}-L_{O3} ))-2C_{44} (10K\nonumber\\
&+6L_1-12L_2+2qL_{O2}+3qL_{O3} ))]+\mathrm{sin}(\varphi)[3C_{11}^2\nonumber\\
&\times(3K+2L_1+L_2+qL_{O1}+2qL_{O3} )+C_{11} (-3\nonumber\\
&\times C_{12}(6K+L_1-L_2+2L_3+2qL_{O1}+3qL_{O2}\nonumber\\
&+qL_{O3} )+10C_{44} (3K+2L_1+L_2+qL_{O1}\nonumber\\
&+2qL_{O3} ))+C_{12} (3C_{12} (3KK-L_1-2L_2+2L_3\nonumber\\
&+qL_{O1}+3qL_{O2}-qL_{O3} )-2C_{44} (15K+9L_1\nonumber\\
&-14L_2+6qL_{O1}+9qL_{O3} ))]\},\label{13}
\end{eqnarray}
\begin{eqnarray}
&\sigma_{33}^{S21}=\frac {M^2}{6C_{11}M_s^2}\mathrm{sin}^2(\varphi)  [-C_{12} L_2+C_{11} (L_1-L_2\nonumber\\
&-qL_{O1}+qL_{O3} )],\nonumber\\
&\sigma_{33}^{S22}=\sigma_{33}^{S23}=\frac {M^2}{24C_kM_s^2}\mathrm{sin}^2(\varphi)\nonumber\\
&\times \{3C_{11}^2 (4L_1-L_2-4qL_{O1}+4qL_{O3} )\nonumber\\
&+C_{11} [-10C_{44} (-4L_1+L_2+4qL_{O1}-4qL_{O3} )\nonumber\\
&+3C_{12} (4L_1-L_2-4L_3-4qL_{O1}+4qL_{O3} )]\nonumber\\
&+2C_{12} [4C_{44} (-3L_1+4L_2+3qL_{O1}-3qL_{O3} )\nonumber\\
&-3C_{12} (4L_1-L_2-2L_3-4qL_{O1}+4qL_{O3} )]\},\label{14}
\end{eqnarray}
\begin{eqnarray}
&\sigma_{33}^{S31}=\sigma_{33}^{S32}=\frac {\mathrm{sin}^2(\varphi)M^2}{6C_{11}M_s^2}\{3C_{11}^2 (K+2L_1-L_2\nonumber\\
&-qL_{O1}+2qL_{O3} )-C_{11} [3C_{12} (2K-L_1+L_2\nonumber\\
&+2L_3+2qL_{O1}+qL_{O2}-qL_{O3} )-10C_{44}(K\nonumber\\
&+2L_1-L_2-qL_{O1}+2qL_{O3} )]+C_{12} [3C_{12} (K\nonumber\\
&-3L_1+2L_2+2L_3+3qL_{O1}+qL_{O2}-3qL_{O3} ) )\nonumber\\
&-2C_{44} (5K+7L_1-2L_2-2qL_{O1}+7qL_{O3} )]\},\nonumber\\
&\sigma_{33}^{S33}=\frac {M^2}{12C_{11}M_s^2}\mathrm{sin}^2(\varphi)[-C_{12} (2K+L_1-5L_2\nonumber\\
&+qL_{O1}+q L_{O3} )+C_{11} (2K+4L_1+L_2-2qL_{O1}\nonumber\\
&+4qL_{O3} )],\label{15}
\end{eqnarray}
and $C_k=3C_{11}^2+10C_{11} C_{44}-3C_{12} (C_{12}+2C_{44} )$. Here, to simplify the formulae, we have set the high order magnetoelastic coefficients $L_{O4}$, $L_{O5}$ and $L_{O6}$ to zero.

Strictly speaking, the free energy is a functional of the magnetization $\bm M$ and the strains $\varepsilon_{ij}$. Due to the magnetoelastic coupling, the elastic fields are related to $\bm M$ at equilibrium state, i.e., the elastic strains $\varepsilon_{ij}=\varepsilon_{ij}({\bm M})$ and the elastic stresses $\sigma_{ij}=\sigma_{ij}({\bm M})$. Thus, $\sigma_{ij}$ and $\varepsilon_{ij}$ have a back-action on $\bm M$ and $\bm M$ should be derived by minimizing $w(\bm M,\varepsilon_{ij} (\bm M))$. In some cases, $\bm M$ can be approximated by $\bm M'$, which is obtained through minimizing $w(\bm M,\varepsilon_{ij}=0)$. The difference between the approximate solution $\bm M'$ and rigorous solution $\bm M$ depends on the magnitude of the relative coefficient $\frac{K^2}{2\alpha_2 (C_{11}+2C_{12})M_s^4}$  \cite {22}. For MnSi, $\frac{K^2}{2\alpha_2 (C_{11}+2C_{12})M_s^4}\approx 10^{-3}$, suggesting that the back-action of strains on the magnetization can be neglected.

As mentioned in section \uppercase\expandafter{\romannumeral 2}, the surface-induced stress field is just the opposite of the skyrmion-induced stress field at the surface, and it fades away as $|z|$ increases. Following the above discussion, such a localized elastic field will also have a back-action on the magnetization $\bm M$.
Generally speaking, the $z$-dependent surface-induced stress field will destroy the 2D structure of the skyrmion lattice and makes it a 3D texture \cite{30,58}. The surface-induced stress field is maximum at the surface, whose magnitude is equivalent to the skyrmion-induced stress field. According to above analysis, the back-action on $\bm M$ is negligible when  $\frac{K^2}{2\alpha_2 (C_{11}+2C_{12})M_s^4}$ is small enough. When  $\frac{K^2}{2\alpha_2 (C_{11}+2C_{12})M_s^4}$ is comparable to 1 (e.g., for materials with strong magnetoelastic coupling), the back-action of the surface-induced stress field on the magnetization has to be taken into account. Instead of solving the exact 3D distribution of $\bm M$, we provide here an approximate method to calculate the effect of this back-action. The exact solution of magnetization $\bm M$ is obtained by minimizing $w(\bm M, \varepsilon_{ij}^{skyrmion} (\bm M)+\varepsilon_{ij}^{surface} (\bm M,z))$, where $\varepsilon_{ij}^{surface} (\bm M,z)$ are the surface-induce elastic strains and $\varepsilon_{ij}^{skyrmion} (\bm M)$ are the skyrmion-induced elastic strains. Since $\varepsilon_{ij}^{surface} (\bm M,z)$ decrease exponentially with $z$, we can overestimate the effect of surface-induced elastic strains by replacing $\varepsilon_{ij}^{surface} (\bm M,z)$ with $\varepsilon_{ij}^{surface} (\bm M,0)$. Minimization of $w(\bm M, \varepsilon_{ij}^{skyrmion} (\bm M)+\varepsilon_{ij}^{surface} (\bm M,0))$ with respect to $\bm M$ yields a 2D magnetization distribution where the back-action of the surface-induced elastic field is considered approximately. 

The discussion of the back-action on the magnetization only applies to internal elastic field but not external. The former one refers to the elastic field induced by $\bm M$ through the magnetoelastic interaction and has a back-action on $\bm M$. The later one is induced by external applied forces or misfit strains, and thus its influence on $\bm M$ is not a back-action. The magnitude of such an influence depends on the strength of the applied external field and usually cannot be ignored. 

\end{appendix}

\bibliographystyle{apsrev4-1}
\bibliography{Skyrmion}

%merlin.mbs apsrev4-1.bst 2010-07-25 4.21a (PWD, AO, DPC) hacked
%Control: key (0)
%Control: author (72) initials jnrlst
%Control: editor formatted (1) identically to author
%Control: production of article title (-1) disabled
%Control: page (0) single
%Control: year (1) truncated
%Control: production of eprint (0) enabled
\begin{thebibliography}{41}%
\makeatletter
\providecommand \@ifxundefined [1]{%
 \@ifx{#1\undefined}
}%
\providecommand \@ifnum [1]{%
 \ifnum #1\expandafter \@firstoftwo
 \else \expandafter \@secondoftwo
 \fi
}%
\providecommand \@ifx [1]{%
 \ifx #1\expandafter \@firstoftwo
 \else \expandafter \@secondoftwo
 \fi
}%
\providecommand \natexlab [1]{#1}%
\providecommand \enquote  [1]{``#1''}%
\providecommand \bibnamefont  [1]{#1}%
\providecommand \bibfnamefont [1]{#1}%
\providecommand \citenamefont [1]{#1}%
\providecommand \href@noop [0]{\@secondoftwo}%
\providecommand \href [0]{\begingroup \@sanitize@url \@href}%
\providecommand \@href[1]{\@@startlink{#1}\@@href}%
\providecommand \@@href[1]{\endgroup#1\@@endlink}%
\providecommand \@sanitize@url [0]{\catcode `\\12\catcode `\$12\catcode
  `\&12\catcode `\#12\catcode `\^12\catcode `\_12\catcode `\%12\relax}%
\providecommand \@@startlink[1]{}%
\providecommand \@@endlink[0]{}%
\providecommand \url  [0]{\begingroup\@sanitize@url \@url }%
\providecommand \@url [1]{\endgroup\@href {#1}{\urlprefix }}%
\providecommand \urlprefix  [0]{URL }%
\providecommand \Eprint [0]{\href }%
\providecommand \doibase [0]{http://dx.doi.org/}%
\providecommand \selectlanguage [0]{\@gobble}%
\providecommand \bibinfo  [0]{\@secondoftwo}%
\providecommand \bibfield  [0]{\@secondoftwo}%
\providecommand \translation [1]{[#1]}%
\providecommand \BibitemOpen [0]{}%
\providecommand \bibitemStop [0]{}%
\providecommand \bibitemNoStop [0]{.\EOS\space}%
\providecommand \EOS [0]{\spacefactor3000\relax}%
\providecommand \BibitemShut  [1]{\csname bibitem#1\endcsname}%
\let\auto@bib@innerbib\@empty
%</preamble>
\bibitem [{\citenamefont {M{\"u}hlbauer}\ \emph {et~al.}(2009)\citenamefont
  {M{\"u}hlbauer}, \citenamefont {Binz}, \citenamefont {Jonietz}, \citenamefont
  {Pfleiderer}, \citenamefont {Rosch}, \citenamefont {Neubauer}, \citenamefont
  {Georgii},\ and\ \citenamefont {B{\"o}ni}}]{1}%
  \BibitemOpen
  \bibfield  {author} {\bibinfo {author} {\bibfnamefont {S.}~\bibnamefont
  {M{\"u}hlbauer}}, \bibinfo {author} {\bibfnamefont {B.}~\bibnamefont {Binz}},
  \bibinfo {author} {\bibfnamefont {F.}~\bibnamefont {Jonietz}}, \bibinfo
  {author} {\bibfnamefont {C.}~\bibnamefont {Pfleiderer}}, \bibinfo {author}
  {\bibfnamefont {A.}~\bibnamefont {Rosch}}, \bibinfo {author} {\bibfnamefont
  {A.}~\bibnamefont {Neubauer}}, \bibinfo {author} {\bibfnamefont
  {R.}~\bibnamefont {Georgii}}, \ and\ \bibinfo {author} {\bibfnamefont
  {P.}~\bibnamefont {B{\"o}ni}},\ }\href@noop {} {\bibfield  {journal}
  {\bibinfo  {journal} {Science}\ }\textbf {\bibinfo {volume} {323}},\ \bibinfo
  {pages} {915} (\bibinfo {year} {2009})}\BibitemShut {NoStop}%
\bibitem [{\citenamefont {Petrova}\ and\ \citenamefont
  {Tchernyshyov}(2011)}]{2}%
  \BibitemOpen
  \bibfield  {author} {\bibinfo {author} {\bibfnamefont {O.}~\bibnamefont
  {Petrova}}\ and\ \bibinfo {author} {\bibfnamefont {O.}~\bibnamefont
  {Tchernyshyov}},\ }\href@noop {} {\bibfield  {journal} {\bibinfo  {journal}
  {Physical Review B}\ }\textbf {\bibinfo {volume} {84}},\ \bibinfo {pages}
  {214433} (\bibinfo {year} {2011})}\BibitemShut {NoStop}%
\bibitem [{\citenamefont {Hu}(2017)}]{59}%
  \BibitemOpen
  \bibfield  {author} {\bibinfo {author} {\bibfnamefont {Y.}~\bibnamefont
  {Hu}},\ }\href@noop {} {\bibfield  {journal} {\bibinfo  {journal}
  {arXiv:1702.01059}\ } (\bibinfo {year} {2017})}\BibitemShut {NoStop}%
\bibitem [{\citenamefont {Bogdanov}\ and\ \citenamefont
  {Yablonskii}(1989)}]{3}%
  \BibitemOpen
  \bibfield  {author} {\bibinfo {author} {\bibfnamefont {A.~N.}\ \bibnamefont
  {Bogdanov}}\ and\ \bibinfo {author} {\bibfnamefont {D.~A.}\ \bibnamefont
  {Yablonskii}},\ }\href@noop {} {\bibfield  {journal} {\bibinfo  {journal}
  {Zh. Eksp. Teor. Fiz}\ }\textbf {\bibinfo {volume} {95}},\ \bibinfo {pages}
  {182} (\bibinfo {year} {1989})}\BibitemShut {NoStop}%
\bibitem [{\citenamefont {Bogdanov}\ and\ \citenamefont {Hubert}(1994)}]{4}%
  \BibitemOpen
  \bibfield  {author} {\bibinfo {author} {\bibfnamefont {A.}~\bibnamefont
  {Bogdanov}}\ and\ \bibinfo {author} {\bibfnamefont {A.}~\bibnamefont
  {Hubert}},\ }\href@noop {} {\bibfield  {journal} {\bibinfo  {journal}
  {Journal of magnetism and magnetic materials}\ }\textbf {\bibinfo {volume}
  {138}},\ \bibinfo {pages} {255} (\bibinfo {year} {1994})}\BibitemShut
  {NoStop}%
\bibitem [{\citenamefont {Moskvin}\ \emph {et~al.}(2013)\citenamefont
  {Moskvin}, \citenamefont {Grigoriev}, \citenamefont {Dyadkin}, \citenamefont
  {Eckerlebe}, \citenamefont {Baenitz}, \citenamefont {Schmidt},\ and\
  \citenamefont {Wilhelm}}]{5}%
  \BibitemOpen
  \bibfield  {author} {\bibinfo {author} {\bibfnamefont {E.}~\bibnamefont
  {Moskvin}}, \bibinfo {author} {\bibfnamefont {S.}~\bibnamefont {Grigoriev}},
  \bibinfo {author} {\bibfnamefont {V.}~\bibnamefont {Dyadkin}}, \bibinfo
  {author} {\bibfnamefont {H.}~\bibnamefont {Eckerlebe}}, \bibinfo {author}
  {\bibfnamefont {M.}~\bibnamefont {Baenitz}}, \bibinfo {author} {\bibfnamefont
  {M.}~\bibnamefont {Schmidt}}, \ and\ \bibinfo {author} {\bibfnamefont
  {H.}~\bibnamefont {Wilhelm}},\ }\href@noop {} {\bibfield  {journal} {\bibinfo
   {journal} {Physical review letters}\ }\textbf {\bibinfo {volume} {110}},\
  \bibinfo {pages} {077207} (\bibinfo {year} {2013})}\BibitemShut {NoStop}%
\bibitem [{\citenamefont {M{\"u}nzer}\ \emph {et~al.}(2010)\citenamefont
  {M{\"u}nzer}, \citenamefont {Neubauer}, \citenamefont {Adams}, \citenamefont
  {M{\"u}hlbauer}, \citenamefont {Franz}, \citenamefont {Jonietz},
  \citenamefont {Georgii}, \citenamefont {B{\"o}ni}, \citenamefont {Pedersen},
  \citenamefont {Schmidt} \emph {et~al.}}]{6}%
  \BibitemOpen
  \bibfield  {author} {\bibinfo {author} {\bibfnamefont {W.}~\bibnamefont
  {M{\"u}nzer}}, \bibinfo {author} {\bibfnamefont {A.}~\bibnamefont
  {Neubauer}}, \bibinfo {author} {\bibfnamefont {T.}~\bibnamefont {Adams}},
  \bibinfo {author} {\bibfnamefont {S.}~\bibnamefont {M{\"u}hlbauer}}, \bibinfo
  {author} {\bibfnamefont {C.}~\bibnamefont {Franz}}, \bibinfo {author}
  {\bibfnamefont {F.}~\bibnamefont {Jonietz}}, \bibinfo {author} {\bibfnamefont
  {R.}~\bibnamefont {Georgii}}, \bibinfo {author} {\bibfnamefont
  {P.}~\bibnamefont {B{\"o}ni}}, \bibinfo {author} {\bibfnamefont
  {B.}~\bibnamefont {Pedersen}}, \bibinfo {author} {\bibfnamefont
  {M.}~\bibnamefont {Schmidt}},  \emph {et~al.},\ }\href@noop {} {\bibfield
  {journal} {\bibinfo  {journal} {Physical Review B}\ }\textbf {\bibinfo
  {volume} {81}},\ \bibinfo {pages} {041203} (\bibinfo {year}
  {2010})}\BibitemShut {NoStop}%
\bibitem [{\citenamefont {Shibata}\ \emph {et~al.}(2013)\citenamefont
  {Shibata}, \citenamefont {Yu}, \citenamefont {Hara}, \citenamefont
  {Morikawa}, \citenamefont {Kanazawa}, \citenamefont {Kimoto}, \citenamefont
  {Ishiwata}, \citenamefont {Matsui},\ and\ \citenamefont {Tokura}}]{7}%
  \BibitemOpen
  \bibfield  {author} {\bibinfo {author} {\bibfnamefont {K.}~\bibnamefont
  {Shibata}}, \bibinfo {author} {\bibfnamefont {X.~Z.}\ \bibnamefont {Yu}},
  \bibinfo {author} {\bibfnamefont {T.}~\bibnamefont {Hara}}, \bibinfo {author}
  {\bibfnamefont {D.}~\bibnamefont {Morikawa}}, \bibinfo {author}
  {\bibfnamefont {N.}~\bibnamefont {Kanazawa}}, \bibinfo {author}
  {\bibfnamefont {K.}~\bibnamefont {Kimoto}}, \bibinfo {author} {\bibfnamefont
  {S.}~\bibnamefont {Ishiwata}}, \bibinfo {author} {\bibfnamefont
  {Y.}~\bibnamefont {Matsui}}, \ and\ \bibinfo {author} {\bibfnamefont
  {Y.}~\bibnamefont {Tokura}},\ }\href@noop {} {\bibfield  {journal} {\bibinfo
  {journal} {Nature nanotechnology}\ }\textbf {\bibinfo {volume} {8}},\
  \bibinfo {pages} {723} (\bibinfo {year} {2013})}\BibitemShut {NoStop}%
\bibitem [{\citenamefont {Tokunaga}\ \emph {et~al.}(2015)\citenamefont
  {Tokunaga}, \citenamefont {Yu}, \citenamefont {White}, \citenamefont
  {R{\o}nnow}, \citenamefont {Morikawa}, \citenamefont {Taguchi},\ and\
  \citenamefont {Tokura}}]{8}%
  \BibitemOpen
  \bibfield  {author} {\bibinfo {author} {\bibfnamefont {Y.}~\bibnamefont
  {Tokunaga}}, \bibinfo {author} {\bibfnamefont {X.~Z.}\ \bibnamefont {Yu}},
  \bibinfo {author} {\bibfnamefont {J.~S.}\ \bibnamefont {White}}, \bibinfo
  {author} {\bibfnamefont {H.~M.}\ \bibnamefont {R{\o}nnow}}, \bibinfo {author}
  {\bibfnamefont {D.}~\bibnamefont {Morikawa}}, \bibinfo {author}
  {\bibfnamefont {Y.}~\bibnamefont {Taguchi}}, \ and\ \bibinfo {author}
  {\bibfnamefont {Y.}~\bibnamefont {Tokura}},\ }\href@noop {} {\bibfield
  {journal} {\bibinfo  {journal} {Nature communications}\ }\textbf {\bibinfo
  {volume} {6}},\ \bibinfo {pages} {7638} (\bibinfo {year} {2015})}\BibitemShut
  {NoStop}%
\bibitem [{\citenamefont {R{\"o}{\ss}ler}\ \emph {et~al.}(2006)\citenamefont
  {R{\"o}{\ss}ler}, \citenamefont {Bogdanov},\ and\ \citenamefont
  {Pfleiderer}}]{9}%
  \BibitemOpen
  \bibfield  {author} {\bibinfo {author} {\bibfnamefont {U.~K.}\ \bibnamefont
  {R{\"o}{\ss}ler}}, \bibinfo {author} {\bibfnamefont {A.~N.}\ \bibnamefont
  {Bogdanov}}, \ and\ \bibinfo {author} {\bibfnamefont {C.}~\bibnamefont
  {Pfleiderer}},\ }\href@noop {} {\bibfield  {journal} {\bibinfo  {journal}
  {Nature}\ }\textbf {\bibinfo {volume} {442}},\ \bibinfo {pages} {797}
  (\bibinfo {year} {2006})}\BibitemShut {NoStop}%
\bibitem [{\citenamefont {Jonietz}\ \emph {et~al.}(2010)\citenamefont
  {Jonietz}, \citenamefont {M{\"u}hlbauer}, \citenamefont {Pfleiderer},
  \citenamefont {Neubauer}, \citenamefont {M{\"u}nzer}, \citenamefont {Bauer},
  \citenamefont {Adams}, \citenamefont {Georgii}, \citenamefont {B{\"o}ni},
  \citenamefont {Duine} \emph {et~al.}}]{10}%
  \BibitemOpen
  \bibfield  {author} {\bibinfo {author} {\bibfnamefont {F.}~\bibnamefont
  {Jonietz}}, \bibinfo {author} {\bibfnamefont {S.}~\bibnamefont
  {M{\"u}hlbauer}}, \bibinfo {author} {\bibfnamefont {C.}~\bibnamefont
  {Pfleiderer}}, \bibinfo {author} {\bibfnamefont {A.}~\bibnamefont
  {Neubauer}}, \bibinfo {author} {\bibfnamefont {W.}~\bibnamefont
  {M{\"u}nzer}}, \bibinfo {author} {\bibfnamefont {A.}~\bibnamefont {Bauer}},
  \bibinfo {author} {\bibfnamefont {T.}~\bibnamefont {Adams}}, \bibinfo
  {author} {\bibfnamefont {R.}~\bibnamefont {Georgii}}, \bibinfo {author}
  {\bibfnamefont {P.}~\bibnamefont {B{\"o}ni}}, \bibinfo {author}
  {\bibfnamefont {R.}~\bibnamefont {Duine}},  \emph {et~al.},\ }\href@noop {}
  {\bibfield  {journal} {\bibinfo  {journal} {Science}\ }\textbf {\bibinfo
  {volume} {330}},\ \bibinfo {pages} {1648} (\bibinfo {year}
  {2010})}\BibitemShut {NoStop}%
\bibitem [{\citenamefont {Seki}\ \emph {et~al.}(2012)\citenamefont {Seki},
  \citenamefont {Ishiwata},\ and\ \citenamefont {Tokura}}]{11}%
  \BibitemOpen
  \bibfield  {author} {\bibinfo {author} {\bibfnamefont {S.}~\bibnamefont
  {Seki}}, \bibinfo {author} {\bibfnamefont {S.}~\bibnamefont {Ishiwata}}, \
  and\ \bibinfo {author} {\bibfnamefont {Y.}~\bibnamefont {Tokura}},\
  }\href@noop {} {\bibfield  {journal} {\bibinfo  {journal} {Physical Review
  B}\ }\textbf {\bibinfo {volume} {86}},\ \bibinfo {pages} {060403} (\bibinfo
  {year} {2012})}\BibitemShut {NoStop}%
\bibitem [{\citenamefont {Kiselev}\ \emph {et~al.}(2011)\citenamefont
  {Kiselev}, \citenamefont {Bogdanov}, \citenamefont {Sch{\"a}fer},\ and\
  \citenamefont {R{\"o}{\ss}ler}}]{12}%
  \BibitemOpen
  \bibfield  {author} {\bibinfo {author} {\bibfnamefont {N.~S.}\ \bibnamefont
  {Kiselev}}, \bibinfo {author} {\bibfnamefont {A.~N.}\ \bibnamefont
  {Bogdanov}}, \bibinfo {author} {\bibfnamefont {R.}~\bibnamefont
  {Sch{\"a}fer}}, \ and\ \bibinfo {author} {\bibfnamefont {U.~K.}\ \bibnamefont
  {R{\"o}{\ss}ler}},\ }\href@noop {} {\bibfield  {journal} {\bibinfo  {journal}
  {Journal of Physics D: Applied Physics}\ }\textbf {\bibinfo {volume} {44}},\
  \bibinfo {pages} {392001} (\bibinfo {year} {2011})}\BibitemShut {NoStop}%
\bibitem [{\citenamefont {Kosevich}\ \emph {et~al.}(1981)\citenamefont
  {Kosevich}, \citenamefont {Ivanov},\ and\ \citenamefont {Kovalev}}]{51}%
  \BibitemOpen
  \bibfield  {author} {\bibinfo {author} {\bibfnamefont {A.~M.}\ \bibnamefont
  {Kosevich}}, \bibinfo {author} {\bibfnamefont {B.~A.}\ \bibnamefont
  {Ivanov}}, \ and\ \bibinfo {author} {\bibfnamefont {A.~S.}\ \bibnamefont
  {Kovalev}},\ }\href@noop {} {\bibfield  {journal} {\bibinfo  {journal}
  {Physica D: Nonlinear Phenomena}\ }\textbf {\bibinfo {volume} {3}},\ \bibinfo
  {pages} {363} (\bibinfo {year} {1981})}\BibitemShut {NoStop}%
\bibitem [{\citenamefont {Kosevich}\ \emph {et~al.}(1990)\citenamefont
  {Kosevich}, \citenamefont {Ivanov},\ and\ \citenamefont {Kovalev}}]{52}%
  \BibitemOpen
  \bibfield  {author} {\bibinfo {author} {\bibfnamefont {A.~M.}\ \bibnamefont
  {Kosevich}}, \bibinfo {author} {\bibfnamefont {B.~A.}\ \bibnamefont
  {Ivanov}}, \ and\ \bibinfo {author} {\bibfnamefont {A.~S.}\ \bibnamefont
  {Kovalev}},\ }\href@noop {} {\bibfield  {journal} {\bibinfo  {journal}
  {Physics Reports}\ }\textbf {\bibinfo {volume} {194}},\ \bibinfo {pages}
  {117} (\bibinfo {year} {1990})}\BibitemShut {NoStop}%
\bibitem [{\citenamefont {Onose}\ \emph {et~al.}(2012)\citenamefont {Onose},
  \citenamefont {Okamura}, \citenamefont {Seki}, \citenamefont {Ishiwata},\
  and\ \citenamefont {Tokura}}]{15}%
  \BibitemOpen
  \bibfield  {author} {\bibinfo {author} {\bibfnamefont {Y.}~\bibnamefont
  {Onose}}, \bibinfo {author} {\bibfnamefont {Y.}~\bibnamefont {Okamura}},
  \bibinfo {author} {\bibfnamefont {S.}~\bibnamefont {Seki}}, \bibinfo {author}
  {\bibfnamefont {S.}~\bibnamefont {Ishiwata}}, \ and\ \bibinfo {author}
  {\bibfnamefont {Y.}~\bibnamefont {Tokura}},\ }\href@noop {} {\bibfield
  {journal} {\bibinfo  {journal} {Physical review letters}\ }\textbf {\bibinfo
  {volume} {109}},\ \bibinfo {pages} {037603} (\bibinfo {year}
  {2012})}\BibitemShut {NoStop}%
\bibitem [{\citenamefont {Yu}\ \emph {et~al.}(2015)\citenamefont {Yu},
  \citenamefont {Kikkawa}, \citenamefont {Morikawa}, \citenamefont {Shibata},
  \citenamefont {Tokunaga}, \citenamefont {Taguchi},\ and\ \citenamefont
  {Tokura}}]{14}%
  \BibitemOpen
  \bibfield  {author} {\bibinfo {author} {\bibfnamefont {X.}~\bibnamefont
  {Yu}}, \bibinfo {author} {\bibfnamefont {A.}~\bibnamefont {Kikkawa}},
  \bibinfo {author} {\bibfnamefont {D.}~\bibnamefont {Morikawa}}, \bibinfo
  {author} {\bibfnamefont {K.}~\bibnamefont {Shibata}}, \bibinfo {author}
  {\bibfnamefont {Y.}~\bibnamefont {Tokunaga}}, \bibinfo {author}
  {\bibfnamefont {Y.}~\bibnamefont {Taguchi}}, \ and\ \bibinfo {author}
  {\bibfnamefont {Y.}~\bibnamefont {Tokura}},\ }\href@noop {} {\bibfield
  {journal} {\bibinfo  {journal} {Physical Review B}\ }\textbf {\bibinfo
  {volume} {91}},\ \bibinfo {pages} {054411} (\bibinfo {year}
  {2015})}\BibitemShut {NoStop}%
\bibitem [{\citenamefont {Nii}\ \emph {et~al.}(2015)\citenamefont {Nii},
  \citenamefont {Nakajima}, \citenamefont {Kikkawa}, \citenamefont {Yamasaki},
  \citenamefont {Ohishi}, \citenamefont {Suzuki}, \citenamefont {Taguchi},
  \citenamefont {Arima}, \citenamefont {Tokura},\ and\ \citenamefont
  {Iwasa}}]{16}%
  \BibitemOpen
  \bibfield  {author} {\bibinfo {author} {\bibfnamefont {Y.}~\bibnamefont
  {Nii}}, \bibinfo {author} {\bibfnamefont {T.}~\bibnamefont {Nakajima}},
  \bibinfo {author} {\bibfnamefont {A.}~\bibnamefont {Kikkawa}}, \bibinfo
  {author} {\bibfnamefont {Y.}~\bibnamefont {Yamasaki}}, \bibinfo {author}
  {\bibfnamefont {K.}~\bibnamefont {Ohishi}}, \bibinfo {author} {\bibfnamefont
  {J.}~\bibnamefont {Suzuki}}, \bibinfo {author} {\bibfnamefont
  {Y.}~\bibnamefont {Taguchi}}, \bibinfo {author} {\bibfnamefont
  {T.}~\bibnamefont {Arima}}, \bibinfo {author} {\bibfnamefont
  {Y.}~\bibnamefont {Tokura}}, \ and\ \bibinfo {author} {\bibfnamefont
  {Y.}~\bibnamefont {Iwasa}},\ }\href@noop {} {\bibfield  {journal} {\bibinfo
  {journal} {Nature communications}\ }\textbf {\bibinfo {volume} {6}},\
  \bibinfo {pages} {8539} (\bibinfo {year} {2015})}\BibitemShut {NoStop}%
\bibitem [{\citenamefont {Chacon}\ \emph {et~al.}(2015)\citenamefont {Chacon},
  \citenamefont {Bauer}, \citenamefont {Adams}, \citenamefont {Rucker},
  \citenamefont {Brandl}, \citenamefont {Georgii}, \citenamefont {Garst},\ and\
  \citenamefont {Pfleiderer}}]{17}%
  \BibitemOpen
  \bibfield  {author} {\bibinfo {author} {\bibfnamefont {A.}~\bibnamefont
  {Chacon}}, \bibinfo {author} {\bibfnamefont {A.}~\bibnamefont {Bauer}},
  \bibinfo {author} {\bibfnamefont {T.}~\bibnamefont {Adams}}, \bibinfo
  {author} {\bibfnamefont {F.}~\bibnamefont {Rucker}}, \bibinfo {author}
  {\bibfnamefont {G.}~\bibnamefont {Brandl}}, \bibinfo {author} {\bibfnamefont
  {R.}~\bibnamefont {Georgii}}, \bibinfo {author} {\bibfnamefont
  {M.}~\bibnamefont {Garst}}, \ and\ \bibinfo {author} {\bibfnamefont
  {C.}~\bibnamefont {Pfleiderer}},\ }\href@noop {} {\bibfield  {journal}
  {\bibinfo  {journal} {Physical review letters}\ }\textbf {\bibinfo {volume}
  {115}},\ \bibinfo {pages} {267202} (\bibinfo {year} {2015})}\BibitemShut
  {NoStop}%
\bibitem [{\citenamefont {Nii}\ \emph {et~al.}(2014)\citenamefont {Nii},
  \citenamefont {Kikkawa}, \citenamefont {Taguchi}, \citenamefont {Tokura},\
  and\ \citenamefont {Iwasa}}]{18}%
  \BibitemOpen
  \bibfield  {author} {\bibinfo {author} {\bibfnamefont {Y.}~\bibnamefont
  {Nii}}, \bibinfo {author} {\bibfnamefont {A.}~\bibnamefont {Kikkawa}},
  \bibinfo {author} {\bibfnamefont {Y.}~\bibnamefont {Taguchi}}, \bibinfo
  {author} {\bibfnamefont {Y.}~\bibnamefont {Tokura}}, \ and\ \bibinfo {author}
  {\bibfnamefont {Y.}~\bibnamefont {Iwasa}},\ }\href@noop {} {\bibfield
  {journal} {\bibinfo  {journal} {Physical review letters}\ }\textbf {\bibinfo
  {volume} {113}},\ \bibinfo {pages} {267203} (\bibinfo {year}
  {2014})}\BibitemShut {NoStop}%
\bibitem [{\citenamefont {Shibata}\ \emph {et~al.}(2015)\citenamefont
  {Shibata}, \citenamefont {Iwasaki}, \citenamefont {Kanazawa}, \citenamefont
  {Aizawa}, \citenamefont {Tanigaki}, \citenamefont {Shirai}, \citenamefont
  {Nakajima}, \citenamefont {Kubota}, \citenamefont {Kawasaki}, \citenamefont
  {Park} \emph {et~al.}}]{19}%
  \BibitemOpen
  \bibfield  {author} {\bibinfo {author} {\bibfnamefont {K.}~\bibnamefont
  {Shibata}}, \bibinfo {author} {\bibfnamefont {J.}~\bibnamefont {Iwasaki}},
  \bibinfo {author} {\bibfnamefont {N.}~\bibnamefont {Kanazawa}}, \bibinfo
  {author} {\bibfnamefont {S.}~\bibnamefont {Aizawa}}, \bibinfo {author}
  {\bibfnamefont {T.}~\bibnamefont {Tanigaki}}, \bibinfo {author}
  {\bibfnamefont {M.}~\bibnamefont {Shirai}}, \bibinfo {author} {\bibfnamefont
  {T.}~\bibnamefont {Nakajima}}, \bibinfo {author} {\bibfnamefont
  {M.}~\bibnamefont {Kubota}}, \bibinfo {author} {\bibfnamefont
  {M.}~\bibnamefont {Kawasaki}}, \bibinfo {author} {\bibfnamefont {H.~S.}\
  \bibnamefont {Park}},  \emph {et~al.},\ }\href@noop {} {\bibfield  {journal}
  {\bibinfo  {journal} {Nature nanotechnology}\ }\textbf {\bibinfo {volume}
  {10}},\ \bibinfo {pages} {589} (\bibinfo {year} {2015})}\BibitemShut
  {NoStop}%
\bibitem [{\citenamefont {Hu}\ and\ \citenamefont
  {Wang}(2016{\natexlab{a}})}]{20}%
  \BibitemOpen
  \bibfield  {author} {\bibinfo {author} {\bibfnamefont {Y.}~\bibnamefont
  {Hu}}\ and\ \bibinfo {author} {\bibfnamefont {B.}~\bibnamefont {Wang}},\
  }\href@noop {} {\bibfield  {journal} {\bibinfo  {journal} {Scientific
  Reports}\ }\textbf {\bibinfo {volume} {6}},\ \bibinfo {pages} {30200}
  (\bibinfo {year} {2016}{\natexlab{a}})}\BibitemShut {NoStop}%
\bibitem [{\citenamefont {Timoshenko}\ and\ \citenamefont
  {Goodier}(1987)}]{24}%
  \BibitemOpen
  \bibfield  {author} {\bibinfo {author} {\bibfnamefont {S.~P.}\ \bibnamefont
  {Timoshenko}}\ and\ \bibinfo {author} {\bibfnamefont {J.~N.}\ \bibnamefont
  {Goodier}},\ }\href@noop {} {\emph {\bibinfo {title} {Theory of
  elasticity}}}\ (\bibinfo  {publisher} {McGraw-Hill},\ \bibinfo {year}
  {1987})\BibitemShut {NoStop}%
\bibitem [{\citenamefont {Wang}(1985)}]{25}%
  \BibitemOpen
  \bibfield  {author} {\bibinfo {author} {\bibfnamefont {M.~Z.}\ \bibnamefont
  {Wang}},\ }\href@noop {} {\bibfield  {journal} {\bibinfo  {journal} {Applied
  Mathematics and Mechanics}\ }\textbf {\bibinfo {volume} {6}},\ \bibinfo
  {pages} {1161} (\bibinfo {year} {1985})}\BibitemShut {NoStop}%
\bibitem [{\citenamefont {Stishov}\ \emph {et~al.}(2008)\citenamefont
  {Stishov}, \citenamefont {Petrova}, \citenamefont {Khasanov}, \citenamefont
  {Panova}, \citenamefont {Shikov}, \citenamefont {Lashley}, \citenamefont
  {Wu},\ and\ \citenamefont {Lograsso}}]{23}%
  \BibitemOpen
  \bibfield  {author} {\bibinfo {author} {\bibfnamefont {S.~M.}\ \bibnamefont
  {Stishov}}, \bibinfo {author} {\bibfnamefont {A.~E.}\ \bibnamefont
  {Petrova}}, \bibinfo {author} {\bibfnamefont {S.}~\bibnamefont {Khasanov}},
  \bibinfo {author} {\bibfnamefont {G.~K.}\ \bibnamefont {Panova}}, \bibinfo
  {author} {\bibfnamefont {A.~A.}\ \bibnamefont {Shikov}}, \bibinfo {author}
  {\bibfnamefont {J.~C.}\ \bibnamefont {Lashley}}, \bibinfo {author}
  {\bibfnamefont {D.}~\bibnamefont {Wu}}, \ and\ \bibinfo {author}
  {\bibfnamefont {T.~A.}\ \bibnamefont {Lograsso}},\ }\href@noop {} {\bibfield
  {journal} {\bibinfo  {journal} {Journal of Experimental and Theoretical
  Physics}\ }\textbf {\bibinfo {volume} {106}},\ \bibinfo {pages} {888}
  (\bibinfo {year} {2008})}\BibitemShut {NoStop}%
\bibitem [{\citenamefont {Hu}\ and\ \citenamefont
  {Wang}(2016{\natexlab{b}})}]{22}%
  \BibitemOpen
  \bibfield  {author} {\bibinfo {author} {\bibfnamefont {Y.}~\bibnamefont
  {Hu}}\ and\ \bibinfo {author} {\bibfnamefont {B.}~\bibnamefont {Wang}},\
  }\href@noop {} {\bibfield  {journal} {\bibinfo  {journal} {arXiv:1604.02766}\
  } (\bibinfo {year} {2016}{\natexlab{b}})}\BibitemShut {NoStop}%
\bibitem [{\citenamefont {Karhu}\ \emph {et~al.}(2012)\citenamefont {Karhu},
  \citenamefont {R{\"o}{\ss}ler}, \citenamefont {Bogdanov}, \citenamefont
  {Kahwaji}, \citenamefont {Kirby}, \citenamefont {Fritzsche}, \citenamefont
  {Robertson}, \citenamefont {Majkrzak},\ and\ \citenamefont {Monchesky}}]{53}%
  \BibitemOpen
  \bibfield  {author} {\bibinfo {author} {\bibfnamefont {E.~A.}\ \bibnamefont
  {Karhu}}, \bibinfo {author} {\bibfnamefont {U.~K.}\ \bibnamefont
  {R{\"o}{\ss}ler}}, \bibinfo {author} {\bibfnamefont {A.~N.}\ \bibnamefont
  {Bogdanov}}, \bibinfo {author} {\bibfnamefont {S.}~\bibnamefont {Kahwaji}},
  \bibinfo {author} {\bibfnamefont {B.~J.}\ \bibnamefont {Kirby}}, \bibinfo
  {author} {\bibfnamefont {H.}~\bibnamefont {Fritzsche}}, \bibinfo {author}
  {\bibfnamefont {M.~D.}\ \bibnamefont {Robertson}}, \bibinfo {author}
  {\bibfnamefont {C.~F.}\ \bibnamefont {Majkrzak}}, \ and\ \bibinfo {author}
  {\bibfnamefont {T.~L.}\ \bibnamefont {Monchesky}},\ }\href@noop {} {\bibfield
   {journal} {\bibinfo  {journal} {Physical Review B}\ }\textbf {\bibinfo
  {volume} {85}},\ \bibinfo {pages} {094429} (\bibinfo {year}
  {2012})}\BibitemShut {NoStop}%
\bibitem [{\citenamefont {Schulz}\ \emph {et~al.}(2012)\citenamefont {Schulz},
  \citenamefont {Ritz}, \citenamefont {Bauer}, \citenamefont {Halder},
  \citenamefont {Wagner}, \citenamefont {Franz}, \citenamefont {Pfleiderer},
  \citenamefont {Everschor}, \citenamefont {Garst},\ and\ \citenamefont
  {Rosch}}]{26}%
  \BibitemOpen
  \bibfield  {author} {\bibinfo {author} {\bibfnamefont {T.}~\bibnamefont
  {Schulz}}, \bibinfo {author} {\bibfnamefont {R.}~\bibnamefont {Ritz}},
  \bibinfo {author} {\bibfnamefont {A.}~\bibnamefont {Bauer}}, \bibinfo
  {author} {\bibfnamefont {M.}~\bibnamefont {Halder}}, \bibinfo {author}
  {\bibfnamefont {M.}~\bibnamefont {Wagner}}, \bibinfo {author} {\bibfnamefont
  {C.}~\bibnamefont {Franz}}, \bibinfo {author} {\bibfnamefont
  {C.}~\bibnamefont {Pfleiderer}}, \bibinfo {author} {\bibfnamefont
  {K.}~\bibnamefont {Everschor}}, \bibinfo {author} {\bibfnamefont
  {M.}~\bibnamefont {Garst}}, \ and\ \bibinfo {author} {\bibfnamefont
  {A.}~\bibnamefont {Rosch}},\ }\href@noop {} {\bibfield  {journal} {\bibinfo
  {journal} {Nature Physics}\ }\textbf {\bibinfo {volume} {8}},\ \bibinfo
  {pages} {301} (\bibinfo {year} {2012})}\BibitemShut {NoStop}%
\bibitem [{\citenamefont {Kong}\ and\ \citenamefont {Zang}(2013)}]{27}%
  \BibitemOpen
  \bibfield  {author} {\bibinfo {author} {\bibfnamefont {L.}~\bibnamefont
  {Kong}}\ and\ \bibinfo {author} {\bibfnamefont {J.}~\bibnamefont {Zang}},\
  }\href@noop {} {\bibfield  {journal} {\bibinfo  {journal} {Physical review
  letters}\ }\textbf {\bibinfo {volume} {111}},\ \bibinfo {pages} {067203}
  (\bibinfo {year} {2013})}\BibitemShut {NoStop}%
\bibitem [{\citenamefont {Hu}\ and\ \citenamefont
  {Wang}(2016{\natexlab{c}})}]{28}%
  \BibitemOpen
  \bibfield  {author} {\bibinfo {author} {\bibfnamefont {Y.}~\bibnamefont
  {Hu}}\ and\ \bibinfo {author} {\bibfnamefont {B.}~\bibnamefont {Wang}},\
  }\href@noop {} {\bibfield  {journal} {\bibinfo  {journal} {arXiv:1608.04840}\
  } (\bibinfo {year} {2016}{\natexlab{c}})}\BibitemShut {NoStop}%
\bibitem [{\citenamefont {Bogdanov}\ and\ \citenamefont
  {R{\"o}{\ss}ler}(2001)}]{29}%
  \BibitemOpen
  \bibfield  {author} {\bibinfo {author} {\bibfnamefont {A.~N.}\ \bibnamefont
  {Bogdanov}}\ and\ \bibinfo {author} {\bibfnamefont {U.~K.}\ \bibnamefont
  {R{\"o}{\ss}ler}},\ }\href@noop {} {\bibfield  {journal} {\bibinfo  {journal}
  {Physical review letters}\ }\textbf {\bibinfo {volume} {87}},\ \bibinfo
  {pages} {037203} (\bibinfo {year} {2001})}\BibitemShut {NoStop}%
\bibitem [{\citenamefont {Rybakov}\ \emph {et~al.}(2015)\citenamefont
  {Rybakov}, \citenamefont {Borisov}, \citenamefont {Bl{\"u}gel},\ and\
  \citenamefont {Kiselev}}]{30}%
  \BibitemOpen
  \bibfield  {author} {\bibinfo {author} {\bibfnamefont {F.~N.}\ \bibnamefont
  {Rybakov}}, \bibinfo {author} {\bibfnamefont {A.~B.}\ \bibnamefont
  {Borisov}}, \bibinfo {author} {\bibfnamefont {S.}~\bibnamefont {Bl{\"u}gel}},
  \ and\ \bibinfo {author} {\bibfnamefont {N.~S.}\ \bibnamefont {Kiselev}},\
  }\href@noop {} {\bibfield  {journal} {\bibinfo  {journal} {Physical review
  letters}\ }\textbf {\bibinfo {volume} {115}},\ \bibinfo {pages} {117201}
  (\bibinfo {year} {2015})}\BibitemShut {NoStop}%
\bibitem [{\citenamefont {Heinze}\ \emph {et~al.}(2011)\citenamefont {Heinze},
  \citenamefont {Von~Bergmann}, \citenamefont {Menzel}, \citenamefont {Brede},
  \citenamefont {Kubetzka}, \citenamefont {Wiesendanger}, \citenamefont
  {Bihlmayer},\ and\ \citenamefont {Bl{\"u}gel}}]{31}%
  \BibitemOpen
  \bibfield  {author} {\bibinfo {author} {\bibfnamefont {S.}~\bibnamefont
  {Heinze}}, \bibinfo {author} {\bibfnamefont {K.}~\bibnamefont
  {Von~Bergmann}}, \bibinfo {author} {\bibfnamefont {M.}~\bibnamefont
  {Menzel}}, \bibinfo {author} {\bibfnamefont {J.}~\bibnamefont {Brede}},
  \bibinfo {author} {\bibfnamefont {A.}~\bibnamefont {Kubetzka}}, \bibinfo
  {author} {\bibfnamefont {R.}~\bibnamefont {Wiesendanger}}, \bibinfo {author}
  {\bibfnamefont {G.}~\bibnamefont {Bihlmayer}}, \ and\ \bibinfo {author}
  {\bibfnamefont {S.}~\bibnamefont {Bl{\"u}gel}},\ }\href@noop {} {\bibfield
  {journal} {\bibinfo  {journal} {Nature Physics}\ }\textbf {\bibinfo {volume}
  {7}},\ \bibinfo {pages} {713} (\bibinfo {year} {2011})}\BibitemShut {NoStop}%
\bibitem [{\citenamefont {Ezawa}(2010)}]{55}%
  \BibitemOpen
  \bibfield  {author} {\bibinfo {author} {\bibfnamefont {M.}~\bibnamefont
  {Ezawa}},\ }\href@noop {} {\bibfield  {journal} {\bibinfo  {journal}
  {Physical review letters}\ }\textbf {\bibinfo {volume} {105}},\ \bibinfo
  {pages} {197202} (\bibinfo {year} {2010})}\BibitemShut {NoStop}%
\bibitem [{\citenamefont {Kirakosyan}\ and\ \citenamefont
  {Pokrovsky}(2006)}]{57}%
  \BibitemOpen
  \bibfield  {author} {\bibinfo {author} {\bibfnamefont {A.~S.}\ \bibnamefont
  {Kirakosyan}}\ and\ \bibinfo {author} {\bibfnamefont {V.~L.}\ \bibnamefont
  {Pokrovsky}},\ }\href@noop {} {\bibfield  {journal} {\bibinfo  {journal}
  {Journal of magnetism and magnetic materials}\ }\textbf {\bibinfo {volume}
  {305}},\ \bibinfo {pages} {413} (\bibinfo {year} {2006})}\BibitemShut
  {NoStop}%
\bibitem [{\citenamefont {Abanov}\ and\ \citenamefont {Pokrovsky}(1998)}]{54}%
  \BibitemOpen
  \bibfield  {author} {\bibinfo {author} {\bibfnamefont {A.}~\bibnamefont
  {Abanov}}\ and\ \bibinfo {author} {\bibfnamefont {V.~L.}\ \bibnamefont
  {Pokrovsky}},\ }\href@noop {} {\bibfield  {journal} {\bibinfo  {journal}
  {Physical Review B}\ }\textbf {\bibinfo {volume} {58}},\ \bibinfo {pages}
  {R8889} (\bibinfo {year} {1998})}\BibitemShut {NoStop}%
\bibitem [{\citenamefont {Ivanov}\ and\ \citenamefont
  {Stephanovich}(1989)}]{56}%
  \BibitemOpen
  \bibfield  {author} {\bibinfo {author} {\bibfnamefont {B.~A.}\ \bibnamefont
  {Ivanov}}\ and\ \bibinfo {author} {\bibfnamefont {V.~A.}\ \bibnamefont
  {Stephanovich}},\ }\href@noop {} {\bibfield  {journal} {\bibinfo  {journal}
  {Physics Letters A}\ }\textbf {\bibinfo {volume} {141}},\ \bibinfo {pages}
  {89} (\bibinfo {year} {1989})}\BibitemShut {NoStop}%
\bibitem [{\citenamefont {K{\'e}zsm{\'a}rki}\ \emph {et~al.}(2015)\citenamefont
  {K{\'e}zsm{\'a}rki}, \citenamefont {Bord{\'a}cs}, \citenamefont {Milde},
  \citenamefont {Neuber}, \citenamefont {Eng}, \citenamefont {White},
  \citenamefont {R{\o}nnow}, \citenamefont {Dewhurst}, \citenamefont
  {Mochizuki}, \citenamefont {Yanai} \emph {et~al.}}]{32}%
  \BibitemOpen
  \bibfield  {author} {\bibinfo {author} {\bibfnamefont {I.}~\bibnamefont
  {K{\'e}zsm{\'a}rki}}, \bibinfo {author} {\bibfnamefont {S.}~\bibnamefont
  {Bord{\'a}cs}}, \bibinfo {author} {\bibfnamefont {P.}~\bibnamefont {Milde}},
  \bibinfo {author} {\bibfnamefont {E.}~\bibnamefont {Neuber}}, \bibinfo
  {author} {\bibfnamefont {L.}~\bibnamefont {Eng}}, \bibinfo {author}
  {\bibfnamefont {J.}~\bibnamefont {White}}, \bibinfo {author} {\bibfnamefont
  {H.~M.}\ \bibnamefont {R{\o}nnow}}, \bibinfo {author} {\bibfnamefont
  {C.}~\bibnamefont {Dewhurst}}, \bibinfo {author} {\bibfnamefont
  {M.}~\bibnamefont {Mochizuki}}, \bibinfo {author} {\bibfnamefont
  {K.}~\bibnamefont {Yanai}},  \emph {et~al.},\ }\href@noop {} {\bibfield
  {journal} {\bibinfo  {journal} {Nature materials}\ }\textbf {\bibinfo
  {volume} {14}},\ \bibinfo {pages} {1116} (\bibinfo {year}
  {2015})}\BibitemShut {NoStop}%
\bibitem [{\citenamefont {Huang}\ and\ \citenamefont {Chien}(2012)}]{33}%
  \BibitemOpen
  \bibfield  {author} {\bibinfo {author} {\bibfnamefont {S.~X.}\ \bibnamefont
  {Huang}}\ and\ \bibinfo {author} {\bibfnamefont {C.~L.}\ \bibnamefont
  {Chien}},\ }\href@noop {} {\bibfield  {journal} {\bibinfo  {journal}
  {Physical review letters}\ }\textbf {\bibinfo {volume} {108}},\ \bibinfo
  {pages} {267201} (\bibinfo {year} {2012})}\BibitemShut {NoStop}%
\bibitem [{\citenamefont {Yu}\ \emph {et~al.}(2011)\citenamefont {Yu},
  \citenamefont {Kanazawa}, \citenamefont {Onose}, \citenamefont {Kimoto},
  \citenamefont {Zhang}, \citenamefont {Ishiwata}, \citenamefont {Matsui},\
  and\ \citenamefont {Tokura}}]{34}%
  \BibitemOpen
  \bibfield  {author} {\bibinfo {author} {\bibfnamefont {X.~Z.}\ \bibnamefont
  {Yu}}, \bibinfo {author} {\bibfnamefont {N.}~\bibnamefont {Kanazawa}},
  \bibinfo {author} {\bibfnamefont {Y.}~\bibnamefont {Onose}}, \bibinfo
  {author} {\bibfnamefont {K.}~\bibnamefont {Kimoto}}, \bibinfo {author}
  {\bibfnamefont {W.~Z.}\ \bibnamefont {Zhang}}, \bibinfo {author}
  {\bibfnamefont {S.}~\bibnamefont {Ishiwata}}, \bibinfo {author}
  {\bibfnamefont {Y.}~\bibnamefont {Matsui}}, \ and\ \bibinfo {author}
  {\bibfnamefont {Y.}~\bibnamefont {Tokura}},\ }\href@noop {} {\bibfield
  {journal} {\bibinfo  {journal} {Nature materials}\ }\textbf {\bibinfo
  {volume} {10}},\ \bibinfo {pages} {106} (\bibinfo {year} {2011})}\BibitemShut
  {NoStop}%
\bibitem [{\citenamefont {Rybakov}\ \emph {et~al.}(2016)\citenamefont
  {Rybakov}, \citenamefont {Borisov}, \citenamefont {Bl{\"u}gel},\ and\
  \citenamefont {Kiselev}}]{58}%
  \BibitemOpen
  \bibfield  {author} {\bibinfo {author} {\bibfnamefont {F.~N.}\ \bibnamefont
  {Rybakov}}, \bibinfo {author} {\bibfnamefont {A.~B.}\ \bibnamefont
  {Borisov}}, \bibinfo {author} {\bibfnamefont {S.}~\bibnamefont {Bl{\"u}gel}},
  \ and\ \bibinfo {author} {\bibfnamefont {N.~S.}\ \bibnamefont {Kiselev}},\
  }\href@noop {} {\bibfield  {journal} {\bibinfo  {journal} {New Journal of
  Physics}\ }\textbf {\bibinfo {volume} {18}},\ \bibinfo {pages} {045002}
  (\bibinfo {year} {2016})}\BibitemShut {NoStop}%
\end{thebibliography}%

\end{document}